\documentclass[fleqn,usenatbib]{mnras}
\usepackage{newtxtext,newtxmath}

\usepackage[T1]{fontenc}
\usepackage{ae,aecompl}

\usepackage{multirow}
\usepackage{graphicx}	
\usepackage{amsmath}	
\usepackage{amssymb}	

\newcommand {\apgt} {\ {\raise-.5ex\hbox{$\buildrel>\over\sim$}}\ }

\newcommand {\aplt} {\ {\raise-.5ex\hbox{$\buildrel<\over\sim$}}\ } 

\title[Model Polarization Curves for WR 6]{Polarization Lightcurve Modeling of Corotating Interaction Regions in the Wind of the Wolf-Rayet Star WR 6}

\author[St-Louis et al.]{
N.St-Louis,$^{1}$\thanks{E-mail: stlouis@astro.umontreal.ca}
Patrick Tremblay$^{1}$
Richard Ignace$^{2}$
\\
$^{1}$D\'epartement de Physique, Universit\'e de Montr\'eal, C.P. 6128, Succ. Centre-Ville, Montr\'eal, Qu\'ebec H3C 3J7, Canada\\
$^{2}$Department of Physics \&\ Astronomy, East Tennessee State University, Johnson City, TN\, 37614, USA\\
}

\date{Accepted XXX. Received YYY; in original form ZZZ}

\pubyear{2017}

\begin{document}
\label{firstpage}
\pagerange{\pageref{firstpage}--\pageref{lastpage}}
\maketitle

\begin{abstract}
The intriguing WN4b star WR6 has been known to display epoch-dependent spectroscopic, photometric and polarimetric variability for several decades.  In this paper, we set out to verify if a simplified analytical model in which Corotating Interaction Regions (CIRs) threading an otherwise spherical wind is able to reproduce the many broadband continuum light curves from the literature with a reasonable set of parameters.

We modified the optically thin model we developed in \citet{2015A&A...575A.129I} to approximately account for multiple scattering and used it to fit 13 separate datasets of this star.  By including two CIRs in the wind, we obtained reasonable fits for all datasets with coherent values for the inclination of the rotation axis ($i_0=166^{\circ}$) and for its orientation in the plane of the sky, although in the latter case we obtained two equally acceptable values ($\psi=63^{\circ}$ and $\psi=152^{\circ}$) from the polarimetry.

Additional line profile variation simulations using the Sobolev approximation for the line transfer allowed us to eliminate the $\psi=152^{\circ}$ solution.  With the adopted configuration ($i_0=166^{\circ}$ and $\psi=63^{\circ}$), we were able to reproduce all datasets relatively well with two CIRs located near the stellar equator and always separated by $\sim 90^{\circ}$ in longitude. The epoch-dependency comes from the fact that these CIRs migrate along the surface of the star. Density contrasts smaller than a factor of two and large opening angles for the CIR ($\beta \goa 35^{\circ}$) were found to best reproduce the type of spectroscopic variability reported in the literature. 
\end{abstract}




\section{Introduction}
It has been known for nearly a century that massive stars develop a strong outflow of matter as a result of their high luminosity. The fact that these winds are not always smooth and spherically symmetric is a more recent recognition. Securing ultra high signal-to-noise spectra of two Wolf-Rayet (WR) stars allowed \citet{1988ApJ...334.1038M} to obtain the first direct spectroscopic evidence that these outflows are often  clumpy. The formation of these inhomogeneities are thought to be related to the driving mechanism of the wind, which has been recognized as highly unstable to perturbations \citep{1984ApJ...284..337O,1988ApJ...335..914O,1995ApJ...446..801G}. As for larger scale structures in hot star winds, already in the late seventies, spectropolarimetric observations of the WR star WR6 led \cite{1979ApJ...231L.141M} to conclude that its envelop must be "flattened".

Shortly after, the advent of widespread ultraviolet (UV) satellite observations led to the discovery that discrete narrow absorption components (NACs) are virtually universal in O star P Cygni profiles \citep{1989ApJS...69..527H}. More extensive time series for some O stars showed that such narrow absorption features migrate towards the blue with time and that this is accompanied by 
variations of the blue edge of saturated P Cygni absorption components of strong lines such as C{\sc iv} and N{\sc v} \citep[i.e.][]{1992ApJ...390..266P}. These migrating features became referred to as Discrete Absorption Components (DACs) . It is conceivable that NACs are actually just snapshots of DACs and therefore reflect some fundamental characteristic of radiatively driven winds.

Hydrodynamic simulations by \citet{1996ApJ...462..469C} subsequently showed that a bright perturbation at the surface of a massive star would lead to the development of Corotating Interaction Regions (CIRs) when the high-speed streams generated as a consequence collide with the unperturbed wind and are wrapped around the star by its rotation. This leads to a slowly propagating discontinuity in the radial velocity gradient, which can explain the DACs (and therefore perhaps also the ubiquitous NACs) detected in O-star UV spectra
\citep[e.g.,][]{2001A&A...378..946H,2004A&A...413..959B}.

Since DACs usually cannot be observed in WR spectra because the absorption component of their P Cygni profiles is almost always saturated \citep[exceptions have been found in the UV for WR24 and in the optical for WR134 by][respectively]{1992A&A...266..377P, 2016MNRAS.460.3407A}, evidence for the presence of CIRs in their wind must come from other diagnostics. \citet{2002A&A...395..209D}  investigated the effects of CIRs on optical and near-infrared emission lines. The predicted variability patterns, a characteristic S-shaped feature in greysclale plots of time series of spectra, are very reminiscent of those observed in a few well-observed WR stars such as WR6 \citep{1997ApJ...482..470M} or WR134 \citep{1999ApJ...518..428M}.

WR6 \citep[classified WN4b by][]{1996MNRAS.281..163S} is one of the most extensively observed WR stars showing spectroscopic variability signatures that can be attributed to the presence of CIRs in its wind. Indeed, not only did \citet{1997ApJ...482..470M} find an unambiguous S-shape pattern in optical spectra, but \citet{1995ApJ...452L..57S} also found periodic P Cygni blue edge variations in IUE spectra of this star, very similar to those found associated with DACs in O stars \citep[e.g.][]{1996A&AS..116..257K}. Furthermore, as mentioned above, spectropolarimetric observations have shown that the wind of WR6 is not spherically symmetric. Finally, this star is known to show epoch-dependant periodic variability in photometry and broadband continuum polarimetry \citep[e.g.][]{1992ApJ...397..277R}.

Continuum polarization from massive-star winds is caused by Thomson scattering off the copious free electrons that are present in these ionized outflows. Electron scattering is grey, and therefore broadband continuum polarization measurements obtained in any waveband can be used to characterize the asymmetry at any given time. Time-dependent measurements are even more valuable as they provide the opportunity to obtain more accurate constraints by tracing its structure from different viewing angles as the star rotates.

We have developed a simplified analytical model consisting of a parametric representation of CIRs threading an otherwise
spherical wind \citep{2015A&A...575A.129I}. This model allows for multiple CIRs, represented by spiral-like density enhancements, to be placed at arbitrary latitudes and longitudes on the star. With this model we were able to explore the range of polarimetric light curves that result as the curvature, latitude, and number of CIRs are varied. We did however assume optically thin electron scattering, which is a limitation for it application to a WR star wind.  

The work presented in this paper is twofold. First, our analytical model was modified to take account of multiple scattering in an approximate way. The details of the improvements made to the model are presented in Section 2. The second part consists in applying our model to the many polarization light curves of WR6 already available in the literature. In Section 3, we describe which data were utilized and how they were organized for modelling. In Section 4, the method we employed to fit the various datasets is described and the results discussed. In Section 5, using the Sobolev approximation to produce line profiles, we qualitatively describe the type of line profile variations that are generated by the CIR geometry we have determined from our polarization modelling. Our conclusions can be found in Section 6.

Our goal is to verify if this simplified kinematical description for a CIR is able to qualitatively reproduce all available polarization light curves with a reasonable set of parameters. A reasonable global solution would entail that stellar parameters and viewing angles be common to all datasets while those related to the CIR could apply to only one. 

\section{Model for the Linear Polarization from a CIR}

\subsection{Review of the Optically Thin Limit}

Here we summarize the approach of \citet[hereafter ISP]{2015A&A...575A.129I} for calculation of variable
polarization from CIR-like structures with optically thin scattering, and
describe modifications to accommodate a WR wind that is optically thick
to electron scattering.  Our approach in ISP was twofold:  first, the
adoption of a kinematic description for a CIR structure, based on \citet{2009AJ....137.3339I} and second, modifications to the work of \citet{1977A&A....57..141B} for thin scattering with axisymmetry to the spiral structures of CIRs.

For a CIR that emerges from the photosphere of the star with radius
$R_\ast$ corresponding to where the wind initiates, the equation of
motion for the geometric center of the CIR is given by the following
expression:

\begin{equation}
\phi = \omega\, t + \phi_0 - \frac{R_\ast\sin\,\theta_0}{r_0}\, \left\{
	\frac{1-\xi}{\xi} + b \ln\left[\frac{w}{\xi\,w_0}\right]\right\},
\end{equation}

\noindent where $\phi$ is the azimuth of the local center for the CIR,
$\phi_0$ is a constant, $r$ is the radial distance in the wind, $\theta_0$ is
the latitude at which the CIR emerges from the star, $\xi = R_\ast/r$ and 
$\omega$ is the angular speed of rotation for the star (assumed solid body).
The factor $w$ is the normalized wind velocity, $w=v(r)/v_\infty$,
with $v_\infty$ the wind terminal speed, and the wind velocity given by

\begin{equation}
v(r) = v_\infty\,\left(1-b\,\xi \right).
\end{equation}

\noindent for which the initial wind speed is $w_0 = 1-b$.
Finally, the parameter $r_0 = v_\infty/\omega$ is the 
``winding radius'', which is a length scale associated with how
rapidly the CIR transitions to a spiral pattern.

ISP make a number of simplifying assumptions in the treatment
for the structure of the CIR.  One is that $\theta=\theta_0$ for all
radii.  This means that the geometric center of a CIR resides on the
surface of a cone.  Another is that the CIR has a density that is purely
a function of radius within it bounds.  Another assumption, key for
the evaluation of the variable polarization, is that the cross-section
of the CIR is circular at any radius.  Although the CIR is fundamentally
an asymmetric structure that threads the wind, its cross-section is
axisymmetric with respect to the radial direction for any location of its center.

To evaluate the polarization, ISP employed the results of \citet{1977A&A....57..141B} 
for a differential cross-sectional volume of the CIR,
which being axisymmetric, has an analytic solution.  The polarization
for the full CIR structure amounts to an integration of these cross-sectional
contribution, each one sequentially phase-lagged in accordance with
the above equation of motion.  The approach is detailed in \S~2 of
ISP.

\subsection{Modification for an Optically Thick Wind}

WR winds are optically thick to electron scattering.  In fact, a
pseudo-photosphere forms in the wind itself, so that the atmosphere at
the hydrostatic level is not directly observed.  In the optical band,
this pseudo-photosphere may form, for example, at around $1.5R_\ast$, where $R_\ast$
signifies the base of the wind, corresponding to the location of the hydrostatic star.  For the velocity law from the preceding
section, this distance corresponds to about a third of the wind terminal
speed.

Owing to the effects of multiple scattering, the results of ISP
formally cannot be applied to the case of WR winds.  However,
in an attempt to approximate results expected from a full radiative
transfer calculation, we explore modelling of variable polarization
from CIRs in WR winds by combining the results of ISP with a
``core-halo'' approach.  Here, we are not referring to the traditional 
boundary between the radius at which hydrostatic equilibrium is reached and 
the dynamic stellar wind. Instead, in this approach one determines the location
of the pseudo-photosphere by the condition of optical depth unity along
the line-of-sight.  That radius, hereafter denoted as $R_{\rm ph}
\ge R_\ast$  serves to define a
spherical photosphere.  Everywhere beyond the photosphere is treated as
being optically thin to electron scattering.  The core-halo approach, sometimes
called the ``last scattering approximation'', has been adopted for
a variety of applications, such as in solar physics \citep[e.g.,][]{1994ASSL..189.....S} and resonance scattering polarization for wind lines \citep[e.g.,][]{2004ApJ...609.1018I}.

An important consideration for polarization is the finite disk
depolarization factor.
This factor accounts for the degree of anistropy of the radiation
field at a point of scattering.  \citet{1977A&A....57..141B} considered
a point star of illumination for their results.  Correction
for how a finite star can lead to a more isotropic radiation field,
and therefore depress the degree of polarization for scattering
near the photosphere, was considered by \citet{1987ApJ...317..290C}, and
expanded by \citet{1989ApJ...344..341B}.  Key for this factor is the location
of the photosphere.  

However, the CIR is still assumed to initiate at radius,
$R_\ast$.  There is, of course, wind driving between $R_\ast$ and 
$R_{\rm ph}$, and whatever atmospheric conditions give rise
to the CIR, we assume the equation of motion for its geometric center
from ISP applies.  Consequently, the emergence
of the CIR at $R_{\rm ph}$ must still take into account that the structure
initiates at $R_\ast$, whereas the finite depolarization factor
is expressly related to $R_{\rm ph}$ and not $R_\ast$.

Accommodation for the core-halo approach, with application to
WR winds, is made through the following definitions.  First,
we redefine the inverse normalized radius $\xi = R_\ast/r \rightarrow R_{\rm ph}/r$.
Then the wind velocity law becomes

\begin{equation}
v(r) = v_\infty\,\left( 1 - b_{\rm ph}\,\xi \right),
\end{equation}

\noindent where $b_{\rm ph} = b R_\ast/R_{\rm ph}$, and the
normalized wind speed at the pseudo-photosphere or ``core'' level 
becomes $w_{\rm ph} = 1-b_{\rm ph}$.  Now the solution for the
equation of motion for the geometric center of the CIR becomes

\begin{equation}
\phi = \omega\, t + \phi_0 - \frac{R_{\rm ph}\,\sin\,\theta_0}{r_0}\, \left\{
	\frac{1-\xi}{\xi}+b_{\rm ph} \ln\left[\frac{w}{\xi\,w_{\rm ph}}
	\right]\right\},
\end{equation}

\noindent In this rescaling, the winding radius $r_0$ is a smaller
multiple of $R_{\rm ph}$ than of $R_\ast$, which expresses the idea
that curvature of the CIR develops more rapidly per unit $R_{\rm ph}$
than per unit $R_\ast$.

The location of the pseudo-photosphere is determined by the condition
that the line-of-sight electron scattering optical depth is unity.  This is set by

\begin{equation}
\tau _{\rm es}= 1 = \int_{R_{\rm ph}}^\infty \, n_{\rm e}(r)\,\sigma_T\,dr,
\end{equation}

\noindent where $\sigma_T$ is the Thomson scattering cross-section,
and $n_{\rm e}$ is the number density of electrons in the wind.
Assuming the number of electrons per nucleon is constant throughout
the wind (i.e., the wind ionization is fixed), then the radius
of the pseudo-photosphere is given by

\begin{equation}
R_{\rm ph} = \frac{n_0\, \sigma_T\, R_*^2}{b_{\rm ph}}  \ln \left( \frac
	{1}{1-b_{\rm ph}} \right) ,
\end{equation}

\noindent where $n_0$ is the number density of electrons at the base
of the wind (i.e., at $R_\ast$ where $w=w_0$).

From this point, the calculation of the linear polarization, of $q$ and $u$, and
then the total polarization $p$ and polarization position angle $\psi_P$,
follow ISP, but in the rescaled units of $R_{\rm ph}$.
The functions $\Gamma_{\rm q}$ and $\Gamma_{\rm u}$ in equations (32) and
(33) of ISP have the same form in $\xi$ but are multiplied by
$b_{\rm ph}$ owing to the rescaling.  The integrals themselves are not
the same as for the optically thin case, because the integrands depend
implicitly on the solution $\phi$ only exterior to the pseudo-photosphere,
as the CIR threads the ``halo'' portion of the wind.

The effects of multiple scattering in a proper treatment of the radiative
transfer would undoubtedly alter the quantitative results that are derived
by the core-halo approach just described.  However, one insight that arises 
from the core-halo approximation is that the optical depth parameter,
$\tau_0$, is no longer a free parameter of the model.  For a wind that
is thin to electron scattering down to $R_\ast$, the polarization scales
linearly with $\tau_0$.  The parameter becomes in many cases 
degenerate with other free parameters that influence the amplitude
of the polarization, such as the opening angle of the CIR.
For a WR~wind, the value of $\tau_0 = n_0\,\sigma_T\,R_\ast$ (eq.~[8]
of ISP) can be determined if $R_{\rm ph}$ is specified,
which might be inferred, for example, from the minimum broadening of emission
lines in conjunction with the velocity law.  In such a case, a WR
wind actually removes a free parameter from the ensemble of possibilities
when seeking to match models against data. Alternatively, one can adopt a value of $\tau_0$ from characteristics of the wind ($\dot{M}, {\rm v}_{\infty}$) and the value of the stellar radius $R_{\rm ph}$ (which is directly linked to $R_*$) determined from observations. This is the approach we have adopted in this work.

\section{Timeseries of Polarimetric Observations of WR6 from the Literature}

WR6 is very bright and therefore has been frequently observed using different techniques and in wavelength regions spanning a large fraction of the electromagnetic spectrum. Moreover, it is an intriguing object as it displays periodic but epoch-dependant variations with a period of 3.76 days, without conclusive evidence for a companion \citep{1997ApJ...482..470M}. Broadband polarization is no exception; many datasets in the visible regime have been published and the trend of the variations is known to vary with time \citep[e.g.][]{1992ApJ...397..277R} although the timescale over which this takes place is only roughly known. 

The earliest observations we have found were obtained in the late sixties by \citet{1970ApJ...160.1083S}. In this sparse dataset, no periodicity was detected and the origin of the variations remained unexplained. Almost ten years later, \citet{1980ApJ...236L.149M} secured observations of this star over a 4 month period (January--April 1979) but unfortunately, these authors do not provide more accurate information on the dates and times at which the data were obtained. Securing those observations had been inspired by the report of periodic spectroscopic variations with a period of 3.76 days, first announced by \citet{1979IAUS...83..421F} and interpreted as revealing the presence of a binary companion. The polarimetric variations were indeed strongly modulated on this period and the variations were found to dominate in the second harmonic (i.e. 1.88 days) rather than in the first. A binary interpretation was therefore retained as many other massive binaries presented a similar behaviour.

The bulk of the time series of polarimetric observations we will be modelling in this work are reported in \citet{1989ApJ...343..426D} and \cite{1992ApJ...397..277R} and were obtained between the mid 1980s and the beginning of 1990 at various locations and using  different polarimeters. Although the binary scenario with a low mass companion was also preferred in those two studies, the wealth of data allowed many caveats to this interpretation to be brought to light. \citet{1989ApJ...343..426D} mentions that the shape of the curve remains stable for some time but subsequently changes gradually over a period of a few months, which was recognized as problematic for the binary scenario. \cite{1992ApJ...397..277R} come to a similar conclusion and the introduction of an additional free parameter describing a change in the geometric distribution of the electrons in the WR wind was necessary to explain the variations. Here we will use all available datasets to test the hypothesis that those linear polarization variations are instead caused by the presence of CIRs in the wind of a single WR star. 

All observing runs during which broadband polarimetry was obtained are listed in Table~1.  The observatory at which the data were obtained and the instrument and filter used are provided as well as the range of observing dates. We identified 13 distinct datasets. Since the variations are known to be epoch-dependant, we chose to divide the observations into time intervals rather than by observing runs. \citet{1989ApJ...343..426D} concluded that there was some stability in the shape of the polarization light curve on a period of about two weeks. If individual runs, were shorter that this, we adopted it as one dataset. If it was longer but did not exceed twice this time interval, we also adopted the observing run as one dataset. Data from overlapping observing runs were divided in time interval. The resulting time series have timespans of 10 to 24 days. The only exception to this are the observations which span 4 months obtained in the seventies. For those we have created one single dataset. The datasets we adopted can be found in Table 2 where we list the dates, the observing runs during which the data were obtained, the time interval of the observations and the total number of $q$ and $u$ datapoints included. Formal period searches were carried out by the original authors for all these datasets, except for those of \citet{1970ApJ...160.1083S} and \citet{1980ApJ...236L.149M}, and in all cases a period compatible with the period of 3.766 days of \citet{1986AJ.....91..925L} was found.

\begin{table*}
\begin{center}
\caption{Polarimetric observing runs for WR6}
\vspace{0.3truecm}
\scalebox{0.8}{\begin{tabular}{lccccc}
\hline 
\hline 
No. & Observatory & Instrument & Dates & Filter & Reference \\
\hline
1 & Siding Spring 24 inch  & & January 1966 -- October 1969  & Johnson B & 1\\
2 & Mt Lemmon 40 in & MINIPOL  & January -- April 1979 & Johnson B & 2 \\
3 & Mts Lemmon and Bigelow 40, 60 \&\ 61 in & MINIPOL & 24 October -- 02 November 1985& Blue (Corning 4-96) & 3  \\
4 & Las Campanas 24 in & MINIPOL & 25 February -- 7 April 1986 & Blue (Corning 4-96)  & 3 \\
5 &  Mt Lemmon 40 in & MINIPOL & 7 -- 30 October  1986 & Blue (Corning 4-96)  & 3 \\
6 & SAAO 1m & U. of Cape Town pol. & 3-23 February 1987 & Blue (Corning 4-96)  & 4 \\
7 & La Silla MPI 2.2 m & PISCO & 3--21 October 1988 &  Str\" omgren b  & 5 \\
8 & CASLEO 2.15 m & VATPOL & 13 January -- 1 February 1990 & Blue (narrow) & 5 \\
9 & Las Campanas UTSO 24 in & MINIPOL & 5 February -- 1 March 1990 & G band$^6$ & 5 \\
10 & CASLEO 2.15 m & VATPOL & 19--27 March 1990 & Blue (narrow) & 5 \\
11 & Las Campanas UTSO 24 in & MINIPOL & 17--22 March 1990 & G band$^6$ & 5 \\
12 & SAAO 1m & U. of Cape Town pol. & 19--26 February 1990 & Johnson V & 5 \\
13 & SAAO 1m & U. of Cape Town pol. & 20 March -- 1 April 1990 & Johnson V & 5 \\
\hline
\multicolumn{5}{l}{$^1$\citet{1970ApJ...160.1083S}, $^2$\citet{1980ApJ...236L.149M}, $^3$ \citet{1989ApJ...343..426D}, $^4$\citet{1989ApJ...343..902R}, $^5$\cite{1992ApJ...397..277R}, $^6$ G: $\lambda _c$=5388 \AA , FWHM =352 \AA . }\\
\normalsize
\end{tabular}}
\end{center}
\end{table*}

\begin{table*}
\begin{center}
\caption{Adopted polarization datasets for WR6}
\vspace{0.3truecm}
\begin{tabular}{lcrcc}
\hline 
\hline 
 Dataset& Dates & Observing Runs & Time Interval & Total Number of \\
Label&&&& q and u data points \\
\hline
1 &  30/01/1968 -- 22/03/1969 & 1 & 24 days & 42 \\
2 & 01/1979 -- 04/1979 & 2 & 4 months &  54 \\
3 & 24/10/1985 -- 02/11/1985 & 3 & 10 days & 60 \\
4  & 25/02/1986 -- 10/03/1986 & 4 & 15 days & 70 \\
5  & 11/03/1986 -- 24/03/1986 & 4 & 14 days & 78 \\
6 & 25/03/1986 -- 07/04/1986 &	 4 & 14 days & 64 \\
7 & 7/10/1986 -- 30/10/1986 &  5 & 24 days & 26 \\
8 & 4/02/1987 -- 23/02/1987 & 6 & 20 days & 40 \\
9 & 5/10/1988 -- 18/10/1988 & 7 &14 days & 28 \\
10 & 13/01/1990 -- 28/01/1990 & 8 & 16 days & 46 \\
11 & 29/01/1990 -- 13/02/1990 & 8, 9 & 16 days & 36 \\
12 & 14/02/1990 -- 01/03/1990 & 9,12 & 16 days & 84 \\
13 & 17/03/1990 -- 01 /04/1990 & 10, 11, 13 & 16 days & 62 \\
\hline
\end{tabular}
\end{center}
\end{table*}

Before proceeding with the fits of the observations, we removed the interstellar polarization from the various datasets in order to model only the intrinsic polarization level from the star and its variability. This is generally not an easy task.  The most reliable method is based on the fact that  the polarization produced in the line-of-sight towards the observer is caused by the alignment of dust grains by a magnetic field in the intervening interstellar medium. The resulting polarization has two important characteristics: first, it is constant in time and second, it follows a well known behaviour as a function of wavelength known as the Serkowski law. Following this law, the polarization has a maximum at a wavelength in the optical and diminishes in the ultraviolet and far infrared. \citet{1990ApJ...365L..19S} used this method to determine the interstellar polarization towards WR6 and found a value  at maximum of q$_{\rm IS}$=+0.55\%\ and u$_{\rm IS}$=0\% . In a paper the following year, \citet{1991ApJ...382..301S} carried out a thorough analysis of spectropolarimetric observations of this star, obtained over a period of one month in February-March 1990, in which they studied the behaviour of the polarization in the various emission lines as a function of time. They used this dataset to determine a new measurement for the interstellar polarization and found a value which agreed within the error bars with that of \cite{1990ApJ...365L..19S}. Therefore, we will adopt this value in this work. 

\section{Fitting the Observations}
\subsection{Model parameters}

Our model of polarization by Thomson scattering from CIRs in an otherwise spherically symmetric massive star wind includes many different parameters that must be either set or fitted. It would be a phenomenal task to fit all these parameters. Therefore we have chosen to adopt certain parameters that have been obtained or estimated from observational studies found in the literature. 
\vskip 0.2truecm
\noindent Some parameters are associated with the star itself:
\vskip 0.1truecm
\noindent $\bullet$ {\bf R$_*$} : the hydrostatic radius,\\
\noindent $\bullet$ {\bf P$_{\rm rot}$}: the rotation period,\\
\noindent $\bullet$ {\bf i$_0$}: the inclination of the rotation axis in the line-of-sight,\\
\noindent $\bullet$ {\bf $\psi$}: the orientation of the rotation axis in the plane of the sky (in observer coordinates).
\vskip 0.2truecm
\noindent Other parameters describe the assumed spherically symmetric wind:
\vskip 0.1truecm
\noindent $\bullet$ {\bf $\dot {\rm M}$}: the mass-loss rate,\\
\noindent $\bullet$ {\bf the velocity law}: as stated in Section 2, we  adopt a standard wind velocity law, ${\rm v(r)}={\rm v}_{\infty}(1- {R_* \over r})^{\gamma}$,with\\
\noindent $\bullet$ {\bf v$_{\infty}$}: the terminal velocity of the wind and\\
\noindent $\bullet$ {\bf $\gamma $}: the exponent of the velocity law. For simplicity, we adopt a value of 1 for this parameter \citep[see discussion of \S~2.1 in ][]{1995A&A...299..151H}.\\
\vskip 0.2truecm
\noindent From these parameters, other quantities are calculated:
\vskip 0.1truecm
\noindent $\bullet$ {\bf n$_0$}, the scale of the electron number density of the wind: n$_0$ = ${ {\dot {\rm M}/ \mu _e m_H} /{4\pi R_* ^2 {\rm v}_\infty} }$\\
\noindent $\bullet$ {\bf $\Omega $}, the angular rotation velocity: $\Omega ={2\pi / P}$
\vskip 0.2truecm
\noindent Finally, there are several parameters required to describe each CIR:
\vskip 0.1truecm
\noindent $\bullet$ {\bf $\phi$}, longitude of the CIR,\\
\noindent $\bullet$ {\bf $\theta$}, latitude of the CIR,\\
\noindent $\bullet$ {\bf $\beta$}, half opening angle of the CIR,\\
\noindent $\bullet$ {\bf $\eta$}, density contrast of the CIR, $\eta ={n_{CIR}-n_{sph} \over n_{sph}}$, where $n_{CIR}$ is the number density in the CIR and $n_{sph}$ that of the spherical wind.
\vskip 0.2truecm
\noindent We also define:
\vskip 0.1truecm
\noindent $\bullet$ {\bf $r_0$}, the winding radius of the CIR: $r_0= v_{\infty} / {\Omega}$\\
\noindent $\bullet$ {\bf $\tau _0$}, the optical depth parameter, $\tau _0 = n_0 R_* \sigma _T$,\\
\noindent $\bullet$ {\bf R$_{ph}$}, given by equation (6), the radius at which the optical depth to electron scattering is unity.  For $b=1$ (wind velocity negligible at $R_{*}$ which we adopt in this work), this can be written as R$_{ph} = R_* / (1-e^{-1/ \tau _0 })$

In Table 3 the values of the adopted parameters are listed  as well as the method used to determine them and the associated reference. Also given are the parameters that have been calculated from those adopted parameters. The parameters remaining to be fit are i$_0$, $\psi$ as well as $\phi$ , $\theta$, $\beta$ and $\eta$ for each CIR. However, from ISP, it can easily be seen that the calculated values of $q$ and $u$ depend on the {\em product} of $\eta$ and $\Lambda _0$, which equals $\eta \cos \beta (1-\cos ^2 \beta)$\footnote{Note that equation (21) of \citet{2015A&A...575A.129I} is incorrect. It should have been: $\Lambda =\int _{\mu _0}^1 (1-3\mu ^2)d\mu = -\mu _0 (1-\mu _0^2)\equiv \Lambda _0$.}, indicating clearly that the opening angle and the density contrast in the CIR are indeed indistinguishable. Therefore, we define a new parameter, $\delta =\eta \cos \beta (1-\cos ^2 \beta)$, which we will fit instead.

In ISP,  it was found that winds with one or two CIRs lead to relatively easily identifiable polarization signatures that are reminiscent of the observed linear polarization curves for WR6. By contrast a larger number of structures emerging from a range of azimuth and latitude positions on the star, were found to yield complex polarimetric behaviour. Therefore, the initial fits presented here include two CIRs, as the observed polarization light curves are clearly not characteristic of one CIR only.

\begin{table*}
\begin{center}
\caption{Adopted and calculated parameters for our polarimetric model of CIRs}
\vspace{0.3truecm}
\begin{tabular}{clll}
\hline 
\hline 
Parameter & Method & Value & Reference \\
\hline
R$_*$ &  spectral fitting of &  2.65 R$_{\odot}$ & \citet{2006AandA...457.1015H} \\
&model atmosphere  (PoWR)& & \\
P$_{\rm rot} $& photometry &  3.766 days &\citet[]{1994AJ....107.2179A} \\
${\rm v}_{\infty}$ & Black through of P Cygni absorption & $\sim$1900 km\,s$^{-1}$ &  \citet{1990ApJ...361..607P,1995ApJ...452L..57S}\\
& Near IR line fits & 1820 km\,s$^{-1}$ & \citet{1995AandA...294..529H} \\
& X-Ray line fits & 1950 km\,s$^{-1}$ &  \citet{2015ApJ...815...29H}\\
& {\bf Adopted :} & 1900 km\,s$^{-1}$ &  \\
$\dot {\rm M}$& spectral fitting of   &  &\\
&model atmosphere (PoWR)& 5$\times$ 10$^{-5}$ M$_{\odot}$\, yr$^{-1}$ &   \citet{2006AandA...457.1015H}\\
& spectral fitting of &  &\\
&model atmosphere (CMFGEN) & 2.7 $\times$ 10$^{-5}$ M$_{\odot}$\, yr$^{-1}$ & \citet{2015wrs..conf..357G}\\
& Radio flux fits & 1.9 $\times$ 10$^{-5}$ M$_{\odot}$\, yr$^{-1}$ & \citet{1998AandA...333..956N}\\
$\mu _e$ & Singly to doubly ionized & 2--4 &  \\
n$_0$ & =${\dot {\rm M}/ \mu _e m_H \over  4 \pi R_*^2 {\rm v}_{\infty}}$& &  \\
&From range of $\dot {\rm M}$ and $\mu _e$ & 2.2 -- 11.7 $\times$ 10$^{12}$ cm$^{-2}$ &\\
& {\bf Adopted:}& 4.8 $\times$ 10$^{12}$ cm$^{-2}$&\\
$\Omega$ & =$2\pi /P$&  1.9 $\times$10$^{-5}$ s$^{-1}$&  \\
$\tau _0$ & =${\rm n}_0R_*\sigma _T$ &  0.59 &  \\
R$_{ph}$& =$R_* / (1-e^{-1/\tau _0})$ &  1.22 R$_*$& \\
\hline
\end{tabular}
\end{center}
\end{table*}

\subsection{$\chi ^2$ Fits of Two CIRs}
To fit the data, the Levenberg-Marquardt (LM) nonlinear least-squares minimization algorithm was selected. To do so, we used the LMFIT Python routine. This optimization method uses gradient and curvature information to explore the parameter space. However, it is a local optimizer and will converge to the closest minimum to the initial guess. Therefore, if the parameter space is complex, it is important to choose a good starting point in order to find the global minimum and avoid getting trapped into a local one. As we wished to fit the $q$ and $u$ curves simultaneously because they are linked by the same parameters, we have added the $\chi ^2$ values for each fit and minimized this sum.

Initial exploration of the parameter space revealed that the determination of the various angles was relatively robust; whatever starting values we used always returned very similar solutions for the angles. Resulting solutions from an initial set of fits revealed very similar values of i$_0$ and $\psi$ for fits with the lowest $\chi ^2$ values. As these two parameters should be the same for all datasets because they describe the orientation of the star in the sky, we resolved to adopt an average value of these parameters and redo the fits while keeping these parameters fixed. To calculate the means, we included the top 5-10 solutions of each dataset with a $\chi ^2$ value lower than a certain threshold, which we set individually for each dataset. In total, we used 66 solutions to calculate the means. For i$_0$, the majority of solutions had a value around 165$^{\circ}$ but 29 were around 195$^{\circ}$. Therefore there is a clear ambiguity for the inclination of the stellar axis with respect to the line of sight (either tilted by 15$^{\circ}$ above or below) that we are unable to resolve from these datasets. This is most likely explained by the fact that for a star that is very close to being observed pole-on (as in this particular case) the solutions are degenerate. We chose to adopt the average of the values tilted by 15$^{\circ}$ above the line of sight, 166$\pm$5$^{\circ}$. We note that if we flip the values that are inclined below the line of sight so that they are above, we obtain the exact same average but with a higher standard deviation. For $\psi$, the majority of solutions have a value around 152$^{\circ}$. However, the are six solutions that have a value around 63$^{\circ}$ that have an equally acceptable value of $\chi ^2$, indicating that there is an 90$^{\circ}$ ambiguity for this parameter. The two mean values are 152$\pm$5$^{\circ}$ and 63$\pm$5$^{\circ}$. We choose to retain the two solutions for the time being and will return to this point in Section 5. Note that we have excluded datasets 1 and 2 from the calculation of these means as they are of much poorer quality or have been obtained over a long time period. Adopting the mean value for i$_0$ of 166$^{\circ}$ and the two possible mean values for $\psi$, 152$^{\circ}$ and 63$^{\circ}$ the Levenberg-Marquardt $\chi ^2$ fits where repeated for each dataset. 

The LM fitting technique turned-out to be not quite so robust  for the $\delta$ parameter. In that case very different results were obtained depending on which starting values we chose. To mitigate the problem for this variable, an iterative grid process was adopted. We chose ten different values for each $\delta$ (one per CIR) equally spaced in the 0.1-1 range, creating 100 pairs of $\delta$ values as starting points. For each pair, we then carried out the fits 20 times, retaining only the fit with the smallest $\chi ^2$ for each pair.  The choice of 20 attempts is, of course arbitrary; it was chosen as a compromise between a higher accuracy and a manageable calculation time. The hundred solutions were then examined manually for each dataset and only the best solutions (between $5-10\% $ of the smallest $\chi ^2$ value) were retained.

\subsubsection{Parameter Space}
To have a clearer idea of the nature of the parameter space related to this problem and to help better understand the fit parameters we obtained, we have examined the characteristics of the parameter space of our problem by plotting the values 
of a given parameter as a function of another while all others are kept fixed. Some interesting trends were uncovered, which are illustrated in Figure~1. These colour images present slices of the multi-dimensional parameter space for our adopted stellar inclination of 166$^{\circ}$  and for the two perpendicular values of  $\psi$, which we found to be equally acceptable. The four panels on the left, correspond to $\psi$=152$^{\circ}$ and those on the right  to $\psi$=63$^{\circ}$. In both cases, the rest of the parameters were set to the values corresponding to the solutions of our dataset number 3, that are listed in Table 4. The colours in these plots represent the $\chi ^2$ values coded from black for the smallest to dark red for the highest. In each panel, a yellow dot represents the best solution found by our Levenberg-Marquardt fitting routine. 

First in the top left corner of both parameter spaces we show the $\chi ^2$ values when the stellar inclination, $i_0$ and the orientation of the CIR in the sky, $\psi$, are left to vary. The fact that in polarization an angle is indistinguishable from one rotated by 180$^{\circ}$ is clearly seen as there are two minima on the $\psi$ axis (at 152$^{\circ}$ and 332 $^{\circ}$). For this parameter space, there does not appear to be the same symmetry for the inclination axis suggesting that we can distinguish a star pointing toward from one pointing away from us. However, the fact that there are two maxima on the inclination axis suggest that this might not always be the case. The parameter space corresponding to $\psi$=63$^{\circ}$, displays a very similar behaviour. Note that the scatter on the mean values of these two parameters, i.e. 5$^{\circ}$, can only be considered as a minimum for the uncertainty of the values we obtain for these parameters. To this, we must add a contribution from the width of the minimum in the parameter space, which for dataset 3 can be estimated from Figure 1 to about 5$^{\circ}$ for i$_0$ and 10$^{\circ}$ for $\psi$. Finally, we should also add a contribution from the quality of the final fits (see next section), which would increase these uncertainties further. Therefore, these values should be considered as estimates instead of actual fits.

In the top right panel of each column, we display the values of $\chi ^2$ when $i_0$ and $\phi _1$ the longitude of one of the CIRs is varied. The situation is slightly different for each value of $\psi$. For $\psi$=152$^{\circ}$ on the left, only one minimum is visible in i$_0$ and $\phi$. For $\psi$=63$^{\circ}$, the minimum in $i_0$ appears less defined and the 180$^{\circ}$ ambiguity can now be clearly seen, while on the $\phi$ axis two relatively narrow minima are found separated by 180$^{\circ}$. This 180$^{\circ}$ ambiguity in $\phi$ also appears in the bottom left plot of each column where we show the behaviour of the latitude versus longitude of one of the CIRs. It now becomes clear that this is in fact a hemisphere ambiguity as the 180$^{\circ}$ ambiguity in $\phi$ is associated with a 180$^{\circ} - \theta$ ambiguity, indicating that we are unable to distinguish between a CIR coming out in one hemisphere from another coming out in an opposite direction in the other hemisphere. Finally when the values of $\delta$ of both CIRs are compared in the bottom right plots, we find that when viewing the star with an plane of the sky angle of $\psi$=152$^{\circ}$, one CIR has a delta parameter which is three times higher than the other which indicates that it is either of higher density contrast of it has a larger opening angle. Conversely, if viewing the star with an angle of $\psi$=63$^{\circ}$, the two CIRs have roughly similar $\delta$ values. We note that in this case, the minimum has an elongated shape in the diagonal direction, indicating that several values if $\delta$ pairs are equally acceptable. In the next section, we will see that this is also true for all other datasets with this value of $\psi$.

\begin{figure*}
\centering
\includegraphics[width=0.49\linewidth]{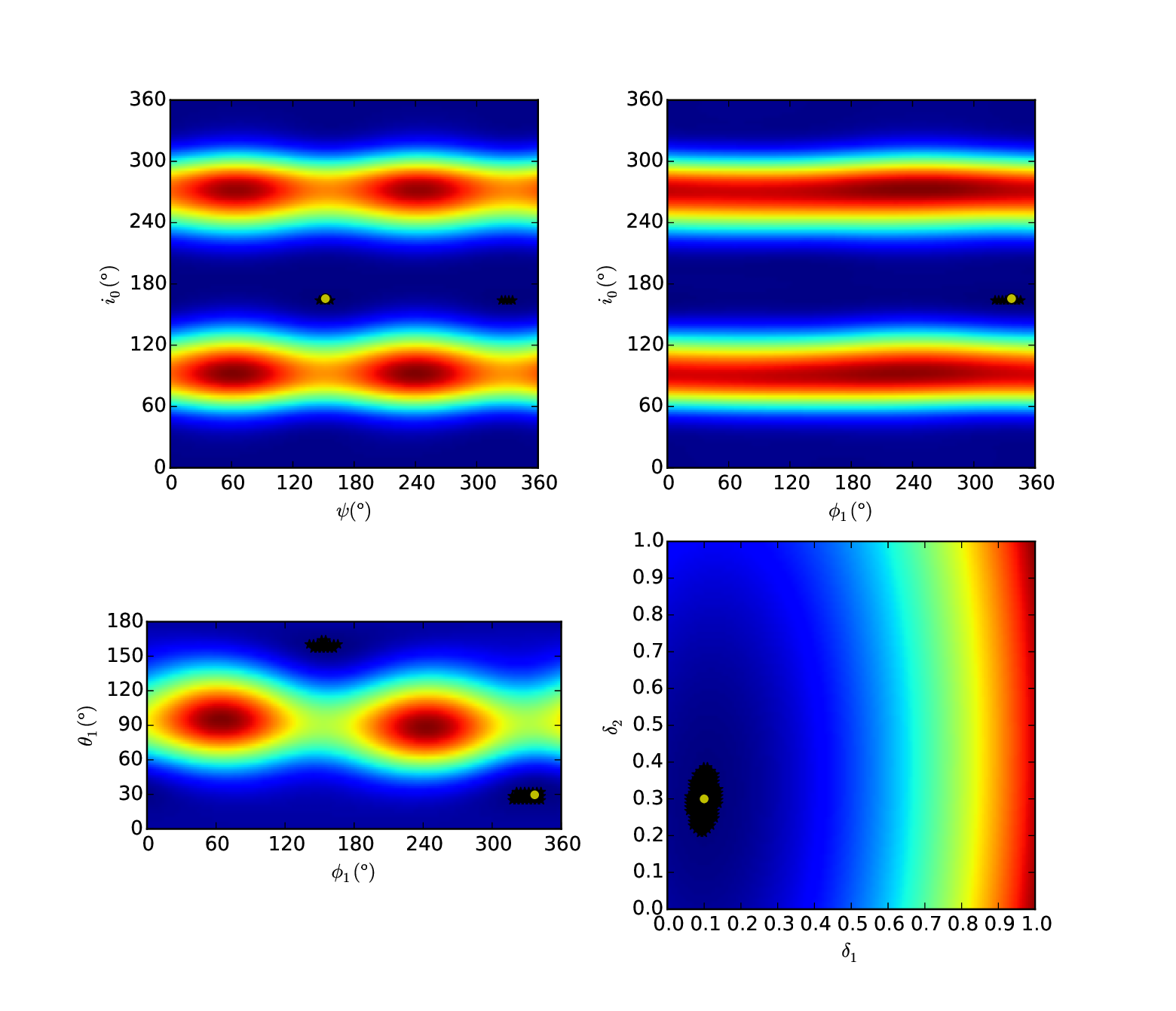}
\hfill
\includegraphics[width=0.49\linewidth]{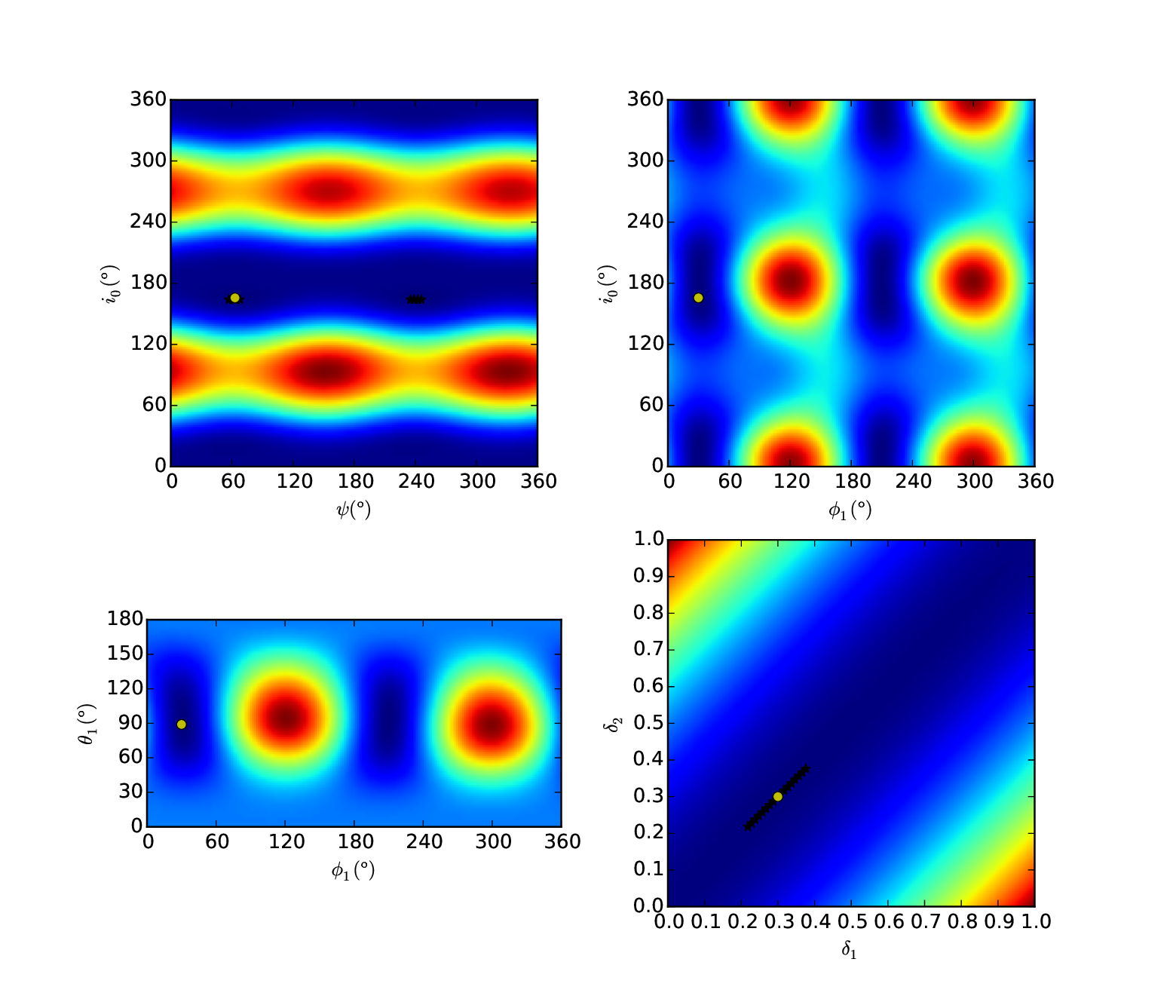}
\caption{Two dimensional slices of the parameter space for an inclination angle of i$_0$=166$^{\circ}$. The four plots in the left column correspond to an angle in the plane of the sky of $\psi$=152$^{\circ}$ and the four plots in the right column correspond to $\psi$=63$^{\circ}$. In each case, the parameters that are fixed for other dimensions are those corresponding to our best fit of dataset number 3. }
\end{figure*}

\subsubsection{Final Fits}
In Table 4,  we list the parameters for the best fit for each available dataset of WR~6 obtained when adopting mean values for $i_0$ and $\psi$. 
For each CIR we list $\theta$, $\phi$ and  $\delta$ as well as the inclination, $i_0$ and value of $\psi$.  The reduced chi square value, $\chi ^2_R$,  listed in this table was obtained by dividing the total $\chi ^2$ for the simultaneous fit of the $q$ and $u$ curves by the degree of freedom defined as the total number of independent data points (given in Table 2) for each dataset minus the number of fitted parameters (six for two CIRs).
Once again, we retain the solutions for the two possible mean values of $\psi$ and will return to this point in the next section.  In reality in each case, in addition the the best solution listed in Table 4, we found a set of solutions that produced only slightly higher values of the reduced $\chi ^2$. The last column of Table 4 gives the number of such solutions for each dataset. The corresponding curves are displayed in Figure 2 using various coloured solid lines superposed on the observational data represented by red dots, for the $\psi$=152$^{\circ}$ case. The curves are visually indistinguishable showing that the fits are equally acceptable.  The main parameter from which these different fits result is the value of $\delta$ or rather the combination of the $\delta$s from each CIR.  Conversely, similar sets of angles  characterizing the position of the CIRs on the star are found. We conclude that even though the $\delta$ values are not well constrained and afflicted with large uncertainties, the angles are much better described. More specifically, it is very difficult to choose which pair of $\delta$s is the correct one; the values we adopted are the ones with the smaller $\chi ^2_R$,  but the others are only marginally worse. For this reason, we provide the range of $\delta$ values that we obtain but specify that each solution corresponds to a specific pair.

\begin{table*}
\begin{center}
\caption{Parameters determined from fits of all datasets with i$_0$ and $\psi$ fixed}
\vspace{0.3truecm}
\begin{tabular}{ccccccccccc}
\hline 
\hline 
Dataset & $\delta _1$ & $\delta _2$ & $\psi$ & i$_0$ & $\phi _1$ & $\phi _2$ & $\theta _1$ & $\theta _2 $& $\chi ^2_R $ &Number of \\
& & & ($^{\circ}$) & ($^{\circ}$) &($^{\circ}$)&($^{\circ}$)&($^{\circ}$)&($^{\circ}$)&  & Solutions \\
\hline
1 & 0.5$^{+0.0}_{-0.1}$ &	1.0$^{+0.0}_{-0.4}$ & 152$\pm$5 & 166$\pm$5 & 269 & 297 & 176 & 7 & 150 & 6 \\
2 & 0.1$^{+0.2}_{-0.0}$  &	0.8$^{+0.0}_{-0.3}$  & 152$\pm$5 & 166$\pm$5 & 330 & 342 & 25 & 178 & 492 & 6 \\
3 & 0.1$^{+0.3}_{-0.0}$ & 0.3$^{+0.6}_{-0.1}$ & 152$\pm$5 & 166$\pm$5 & 337 & 49 & 30 & 175 & 73 & 6 \\
4 & 0.1$^{+0.3}_{-0.0}$ &	0.4$^{+0.3}_{-0.1}$ & 152$\pm$5 & 166$\pm$5 & 129 & 95 & 18 & 176 & 48 & 5 \\
5 & 0.2$^{+0.7}_{-0.0}$ &	0.4$^{+0.4}_{-0.3}$ & 152$\pm$5 & 166$\pm$5 & 119 & 106 & 12 & 177 & 47 & 7 \\
6 & 0.1$^{+0.9}_{-0.0}$ &	0.5$^{+0.1}_{-0.2}$ & 152$\pm$5 & 166$\pm$5 & 105 & 103 & 17 & 178 & 57 & 8 \\
7 & 0.1$^{+0.2}_{-0.0}$ &	0.4$^{+0.3}_{-0.0}$ & 152$\pm$5 & 166$\pm$5 & 246 & 225 & 23 & 176 & 33 & 5 \\
8 & 0.2$^{+0.8}_{-0.1}$ & 0.6$^{+0.1}_{-0.5}$ & 152$\pm$5 & 166$\pm$5 & 150 & 253 & 177 & 4 & 17 & 7 \\
9 & 0.3$^{+0.5}_{-0.0}$ &	0.2$^{+0.7}_{-0.0}$ & 152$\pm$5 & 166$\pm$5 & 249 & 64 & 177 & 173 & 21 & 7 \\
10 & 1.0$^{+0.0}_{-0.6}$ & 0.1$^{+0.9}_{-0.0}$ & 152$\pm$5 & 166$\pm$5 & 46 & 241 & 179 & 25 & 12 & 13 \\
11 & 0.2$^{+0.7}_{-0.1}$ & 0.9$^{+0.0}_{-0.6}$ & 152$\pm$5 & 166$\pm$5 & 257 & 351 & 18 & 179 & 17 & 10\\
12 & 0.8$^{+0.2}_{-0.7}$ & 0.6$^{+0.3}_{-0.5}$ & 152$\pm$5 & 166$\pm$5 & 44 & 72 & 176 & 6 & 34 & 16 \\
13 & 0.4$^{+0.5}_{-0.3}$ & 0.8$^{+0.2}_{-0.7}$ & 152$\pm$5 & 166$\pm$5 & 98 & 56 & 11 & 175 & 51 & 15 \\

\hline
1 & 0.8$^{+0.0}_{-0.1}$ &	0.8$^{+0.0}_{-0.1}$ & 63$\pm$5 & 166$\pm$5 & 268 & 178 & 90 & 89 & 135 & 2 \\
2 & 0.8$^{+0.0}_{-0.1}$  &	0.8$^{+0.0}_{-0.1}$  & 63$\pm$5 & 166$\pm$5 & 210 & 119 & 91 & 90 & 502 & 2 \\
3 & 0.3 &	0.3 & 63$\pm$5 & 166$\pm$5 & 30 & 299 & 89 & 96 & 88 & 1 \\
4 & 0.4 &	0.4 & 63$\pm$5 & 166$\pm$5 & 342 & 71 & 91 & 88 & 48 & 1 \\
5 & 0.4 &	0.4 & 63$\pm$5 & 166$\pm$5 & 330 & 60 & 91 & 89 & 42 & 1 \\
6 & 0.5 &	0.5 & 63$\pm$5 & 166$\pm$5 & 321 & 51 & 91 & 90 & 40 & 1 \\
7 & 0.4 &	0.4 & 63$\pm$5 & 166$\pm$5 & 114 & 24 & 91 & 93 & 36 & 1 \\
8 & 0.3 &	0.3 & 63$\pm$5 & 166$\pm$5 & 159 & 69 & 88 & 88 & 20 & 1 \\
9 & 0.4$^{+0.0}_{-0.1}$  &	0.4$^{+0.0}_{-0.1}$  & 63$\pm$5 & 166$\pm$5 & 25 & 115 & 90 & 90 & 23 & 2 \\
10 & 1.0$^{+0.0}_{-0.1}$  & 1.0$^{+0.0}_{-0.1}$  & 63$\pm$5 & 166$\pm$5 & 86 & 175 & 89 & 91 & 13 & 2 \\
11 & 0.9$^{+0.1}_{-0.0}$  & 0.9$^{+0.1}_{-0.0}$  & 63$\pm$5 & 166$\pm$5 & 228 & 138 & 92 & 90 & 19 & 2 \\
12 & 0.9 & 0.9 & 63$\pm$5 & 166$\pm$5 & 277 & 187 & 92 & 93 & 43 & 1 \\
13 & 0.9$^{+0.0}_{-0.1}$  & 0.9$^{+0.0}_{-0.1}$  & 63$\pm$5 & 166$\pm$5 & 211 & 301 & 93 & 91 & 55 & 2 \\
\hline
\end{tabular}
\end{center}
\end{table*}

\begin{figure*}
\centering
\includegraphics[width=\textwidth]{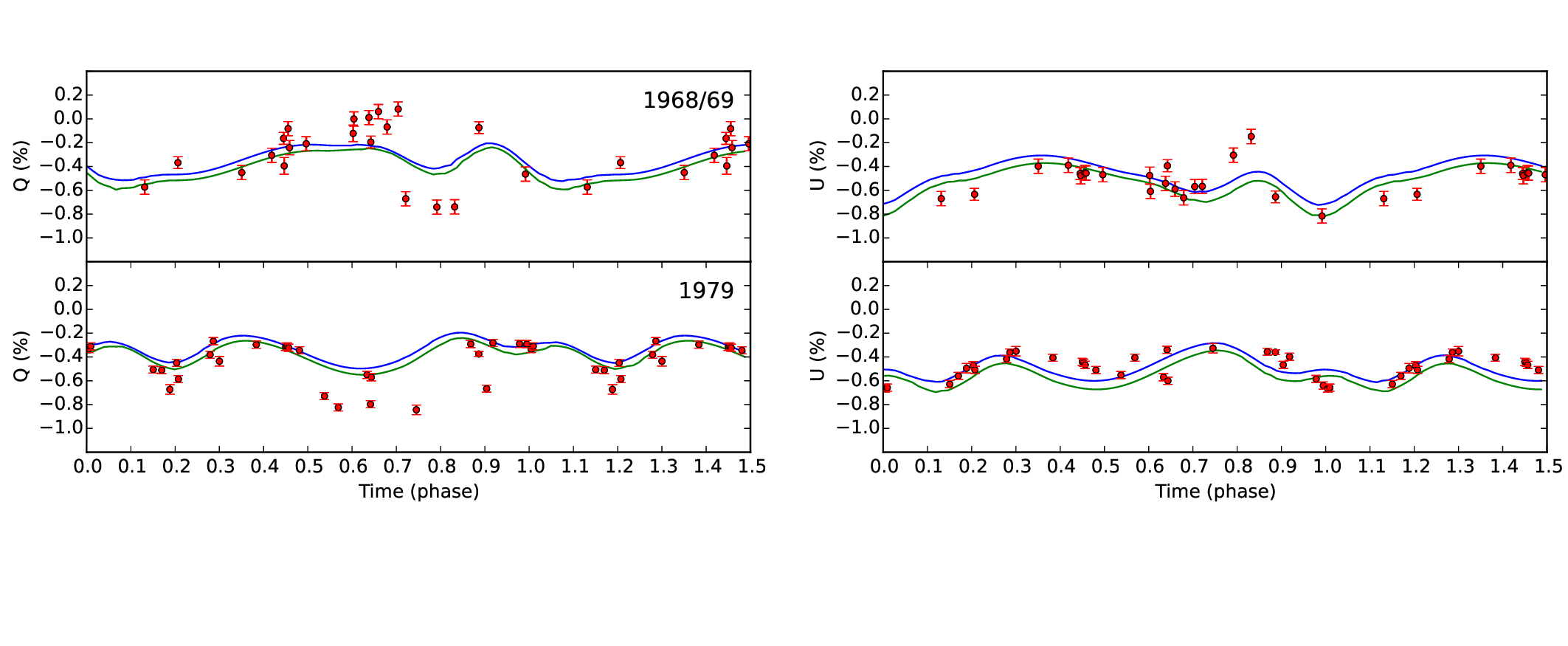}\\
\vskip -1truecm
\includegraphics[width=\textwidth]{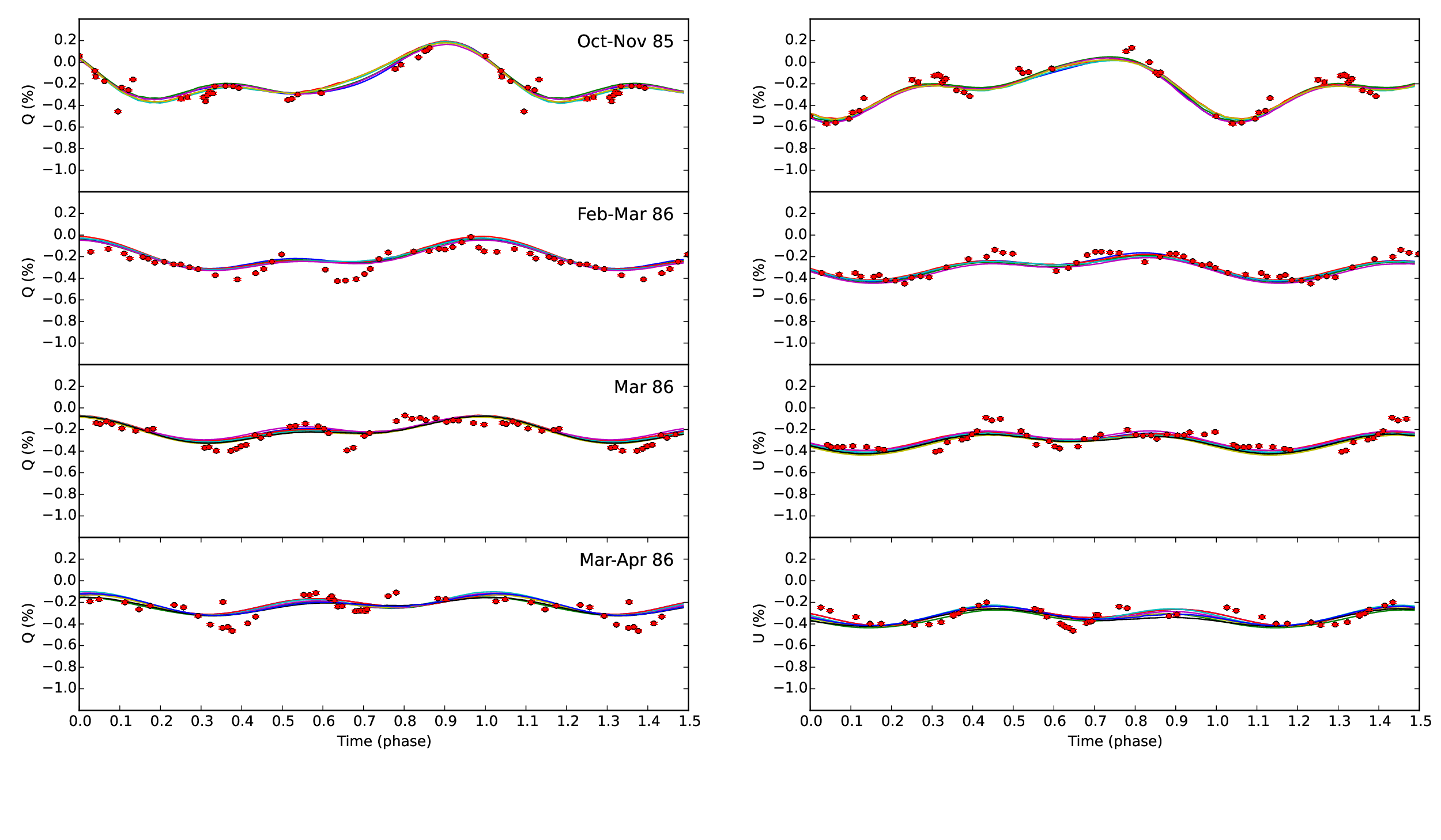}
\vskip -1truecm
\caption{Final fits to all datasets for i$_0$=166$^{\circ}$ and $\psi$=152$^{\circ}$ fixed.  The read dots are the observations and the multi-coloured curves are the equally acceptable fits.}
\end{figure*}

\begin{figure*}
\centering
\includegraphics[width=\textwidth]{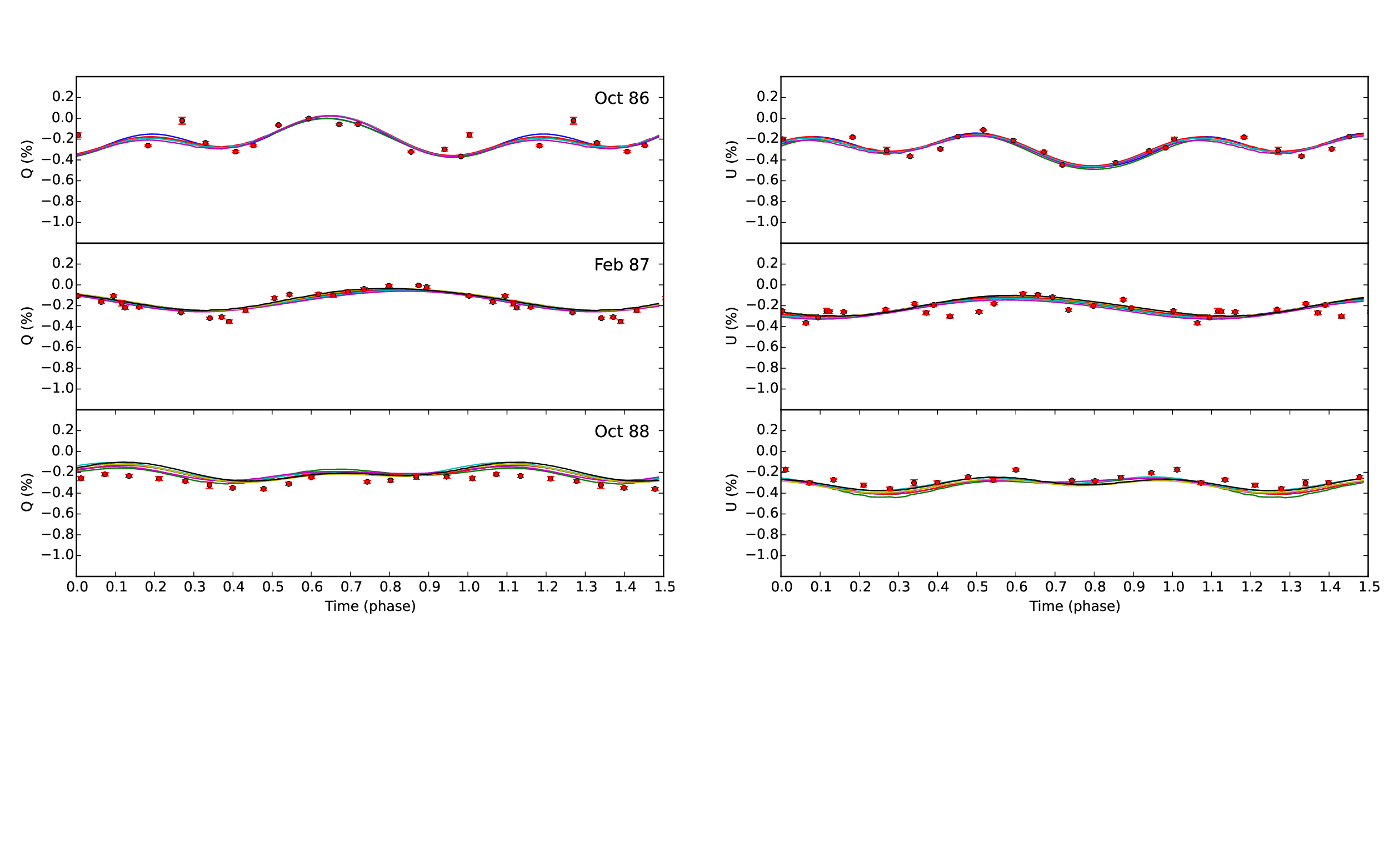}
\vskip -3.5truecm
\includegraphics[width=\textwidth]{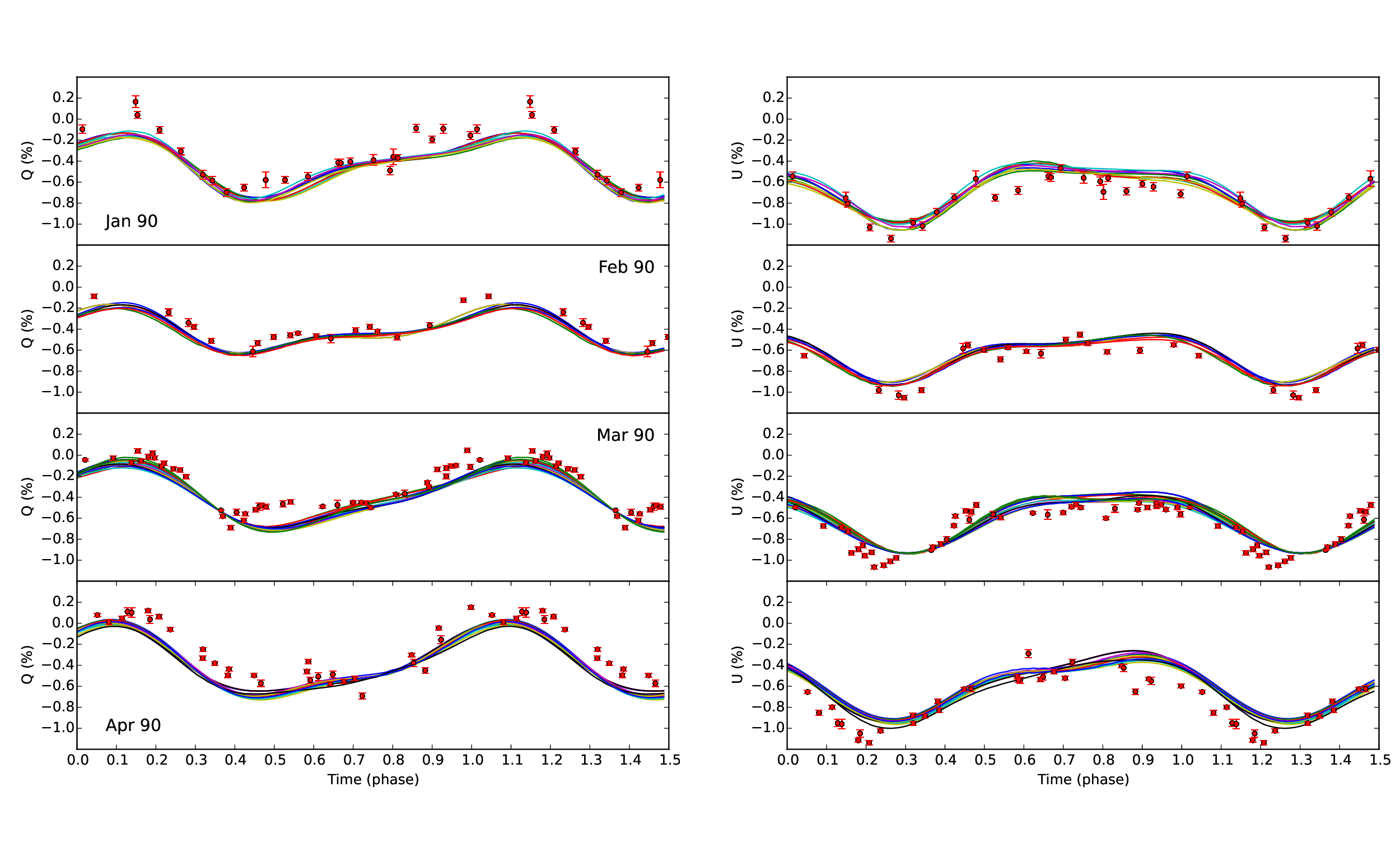}
\vskip -0.5truecm
\contcaption{Final fits to all datasets for i$_0$=166$^{\circ}$ and $\psi$=152$^{\circ}$ fixed.  The read dots are the observations and the multi-coloured curves are the equally acceptable fits.}
\end{figure*}

Recalling that we are fitting simultaneously the mean levels of $q$ and $u$ as well as the shape of both curves, the fits qualitatively reproduce relatively well the observed epoch-dependant periodic polarimetric variations. They are by no means perfect, as reflected by the high values of the $\chi ^2_R$ in all cases but the essence of the variability is reproduced.  Note that the fits to the older data obtained at the end of the 1960s and 1970s are shown more for completeness as we know \citep[e.g.][]{1989ApJ...343..426D} that the shape of the polarization curve changes after a few weeks (although the "stability" timescale has not yet been determined). Therefore, phase-folded curves for data obtained over long periods of time, such as several months or years, are expected to show a scatter higher than the measurement errors because of the varying shape of the curve. In view of this fact, the relatively good agreement of the fits with the data of \citet{1980ApJ...236L.149M} obtained over a four-month period in 1979 is surprising. Even more unexpected is the relatively adequate fit of the U curve to the data of \citet{1970ApJ...160.1083S} obtained over a period of more than three years. This possibly indicates that most of the changes are associated to the Q parameter, i.e. to material located in or perpendicular to the equatorial plane of the star. In the next section, we will examine the possibility of improving the fits by including more than two CIRs at a time.

Examining the fitted parameters in Table 4, an interesting result emerges. Fits with $\psi$=152$^{\circ}$ are mainly associated with CIRs close to the poles ($\theta$=0 or 180$^{\circ}$) while those with  $\psi$=63$^{\circ}$ find CIRs closer to the equator. 
This might seem surprising at first but in fact from the point of view of polarized light, both situations correspond to comparable situations. Indeed for an observer, both geometries correspond to excesses of free electrons with similar distributions on the plane of the sky and therefore generate roughly the same linear polarization from Thomson scattering for an observer (see cartoon in Figure 3). Polarization observations alone are therefore unable to distinguish between these two solutions. For one given value of  $\psi$, the values of the longitudes of the two CIRs are very different from one epoch to another except if the data were obtained over a relatively short timespan (datasets 4, 5 and 6 for example which were obtained over a period of about 40 days).
However, in the case of $\psi$=63$^{\circ}$, for {\em all} epochs they are separated approximately by 90$^{\circ}$, just as if two fixed active regions were slowly drifting along the surface of the star. For $\psi$=152$^{\circ}$, the differences in location in longitude of the two CIRs is more variable but the fact that they so close to the poles renders this parameter much less influential on the resulting polarization curve. Another interesting result for the $\psi$=63$^{\circ}$ configuration is that the values the $\delta$ parameters for the two CIRs is equal in all case, which is not the case for the other configuration. This indicates that in the former case, both CIRs contribute equally to the amplitude of variation of the polarization whereas in the latter case, one of the two CIRs dominates over the other one.

\begin{figure*}
\centering
\vskip -1truecm
\includegraphics[angle=-90,width=\textwidth]{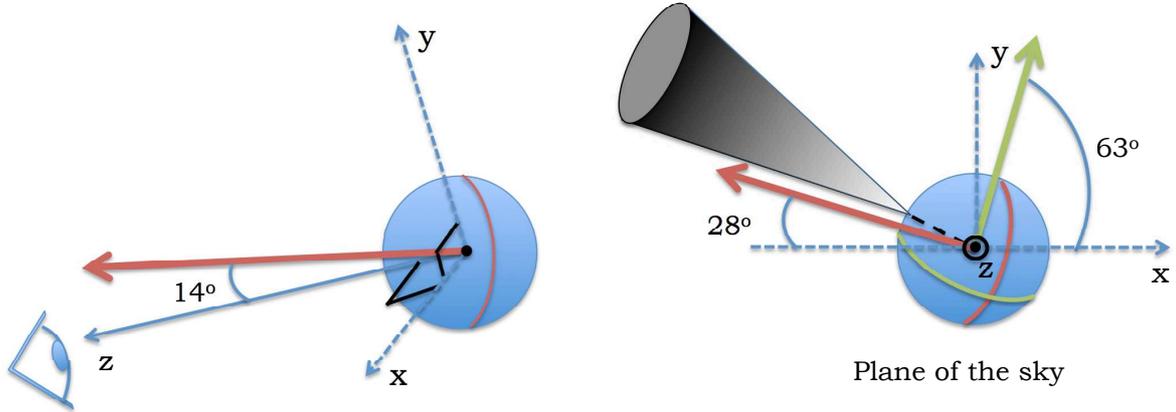}
\vskip -6.3truecm
\caption{Cartoon showing our fitted value of the inclination ($i_0 =166^{\circ}$, left) and the two possible values we obtain for the angle of the stellar axis on the plane of the sky ($\psi =152^{\circ}$ in red and $\psi =63^{\circ}$ in green, right). A CIR density enhancement is schematized by a black cone and illustrates the degeneracy of these two solutions when polarization observations are concerned.}
\end{figure*}

\subsection{$\chi ^2$ Fits of Multiple CIRs}
In the previous section, we performed fits with two CIRs as the non-symmetric shapes of the observed curved indicated that the presence of only one would be insufficient to reproduce the curves. We explored the possibility to improve the fits by adding additional CIRs on the surface of the star. For this test, we have excluded datasets 1 and 2 and have adopted an inclination of 166$\pm$5$^{\circ}$ and $\psi$=152$^{\circ}$ as it was the most commonly found solution.  We then proceeded to add more CIRs one at a time, up to a total of six, to attempt to obtain a better fit to our observations.  In Table~5, we present our results by giving for each dataset the value of  $\chi ^2_R$ value for two CIRs, the optimal number of CIRs found and the new value of  $\chi ^2_R$. Out of 11 datasets, adding more CIRs did not help to secure a better fit. Only in three cases, for datasets 4, 5 and 6, all obtained in a time period of just over 40 days, were we able to obtain a slightly better solution. In the case of dataset 4, the improvement is marginal, as illustrated in Figure 4a , while in for datasets 5 and 6, the fits are significantly improved as can be seen in Figures 4b and 4c and by the values of $\chi ^2_R$ that are considerably reduced.

\begin{table}
\begin{center}
\caption{Fits With Multiple CIRs with i$_0$=166$\pm$5$^{\circ}$ and $\psi$=152$^{\circ}$.}
\begin{tabular}{cccc}
\hline 
\hline 
Dataset&Reduced $\chi ^2$ &Optimal Number&Optimal \\
&for two CIRs&of CIRs&Reduced $\chi ^2$\\
\hline 
3&73&2&73\\
4&48&3&38\\
5&47&6&12\\
6&57&3&30\\
7&33&2&33\\
8&17&2&17\\
9&21&2&21\\
10&12&2&12\\
11&17&2&17\\
12&34&2&34\\
13&51&2&51\\
\hline 
\end{tabular}
\end{center}
\end{table}

\begin{figure}
\centering
\vskip -0.1truecm
\includegraphics[width=0.5\textwidth]{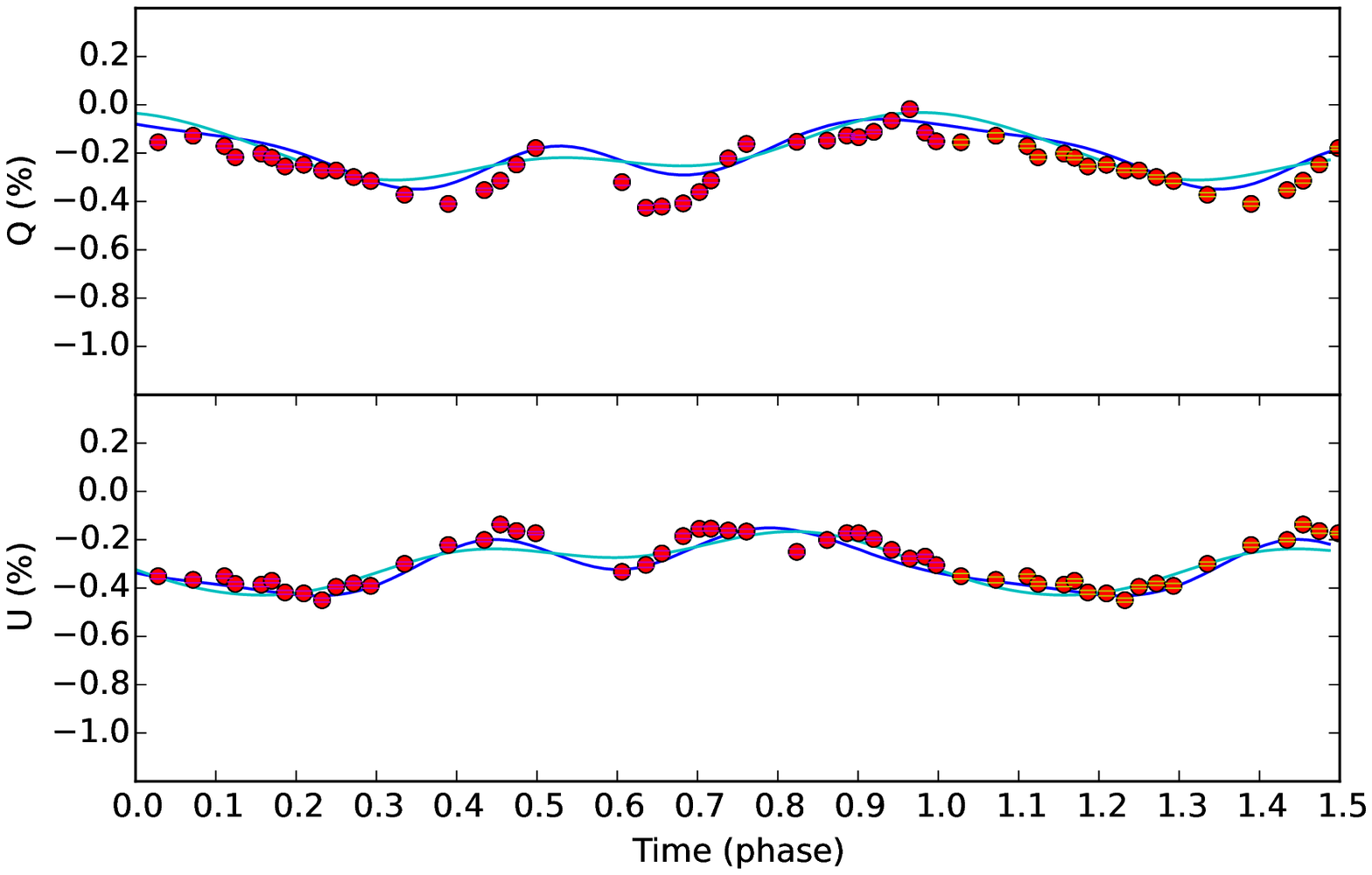}\\
\vskip -0.1truecm
\includegraphics[width=0.5\textwidth]{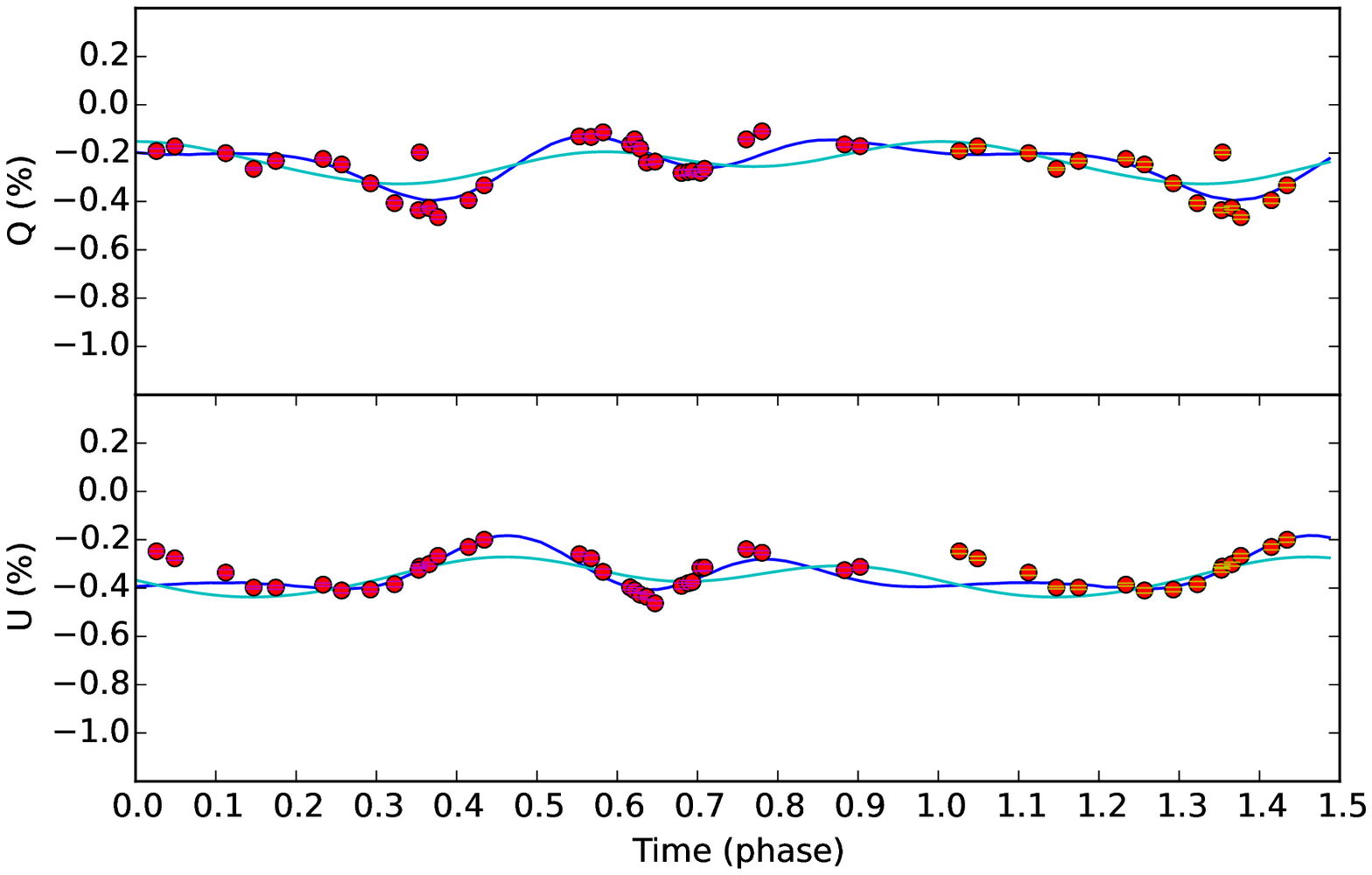}\\
\vskip -0.1truecm
\includegraphics[width=0.5\textwidth]{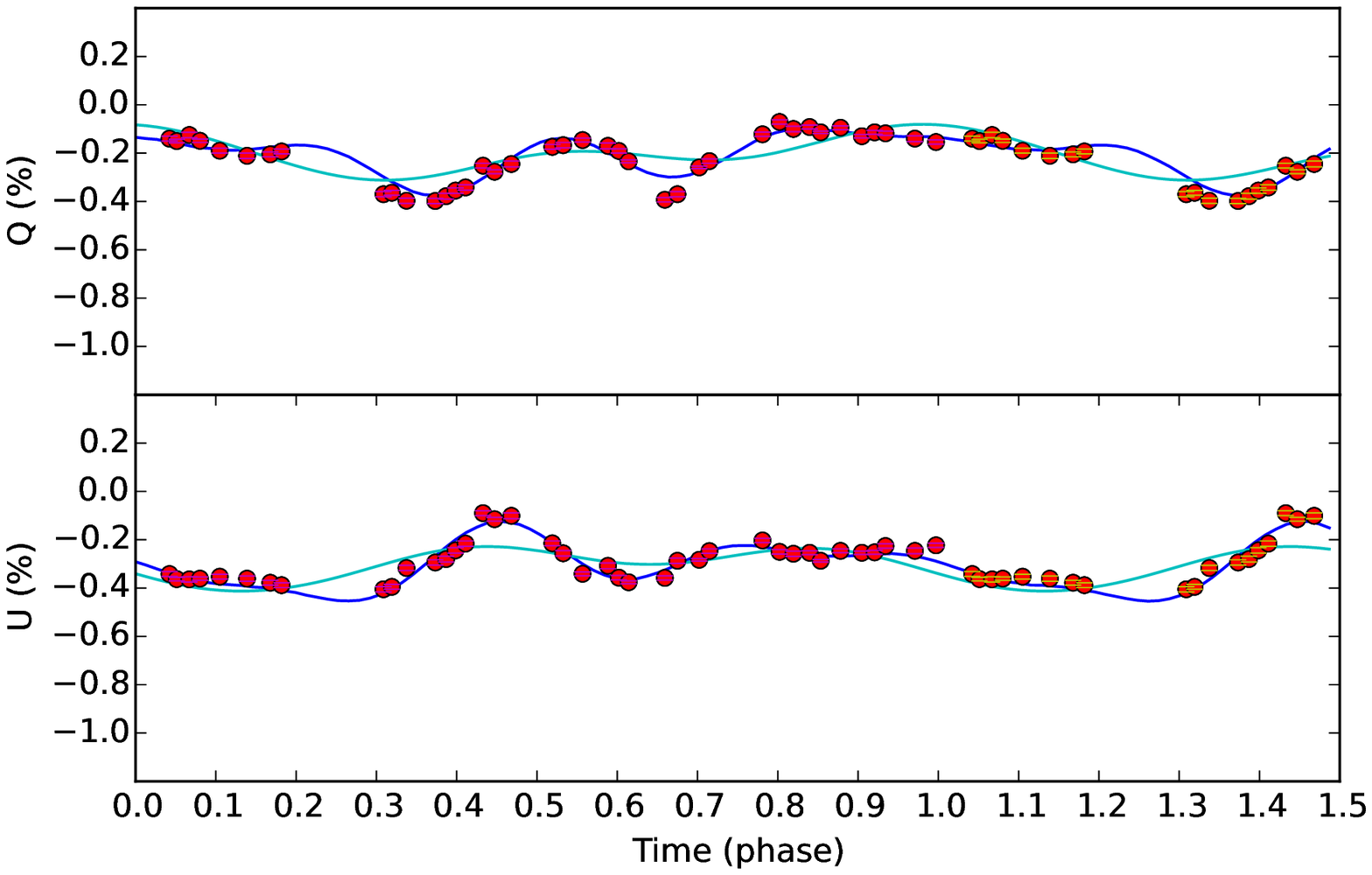}
\caption{Multiple CIR fits. (a) {\it Top} Fit of three CIRs to dataset 4. (b) {\it Middle} Fit of six CIRs to dataset 5.(c) {\it Bottom} Fit of three CIRs to dataset 6.}
\end{figure}

\section{Line Profile Variability}
WR6 is an extremely well observed star. It is well-known to 
present periodic variability not only in polarimetry but also
in spectroscopy. We unfortunately do not have spectroscopic observations 
obtained simultaneously with the polarimetry.  However, in an attempt to ascertain 
if the geometries we have found lead to spectroscopic variability with similar
characteristics as those observed in optical emission lines such
as those presented in \citet{1997ApJ...482..470M}, for example, we set out to calculate simple emission line profiles.
Our goal is not, of course, to carry out time-dependant line profile modelling but simply 
to qualitatively determine the expected spectroscopic variability from our CIR geometries.

 \subsection{Line Profile Model Details}
To explore line profile variability in a quick and qualitative way, we
employed the Sobolev approximation for the line transfer \citep{1960mes..book.....S}.
Given the asymmetry of the outflow, consisting of spiral CIR structures
threading an otherwise spherical wind, we chose to employ the escape
probability approach of  \citet{1983ApJ...274..380R}.  Although generally
one expects a complex velocity field owing to the formation of the CIR
\citep[e.g.,][]{1996ApJ...462..469C}, we have so far treated the CIR as a density
perturbation of the spherical wind, either an excess or deficit in
density scale.  We have assumed that the outflow within the CIR is the
same $\beta=1$ law that applies to the spherical wind.  We have also
assumed the same terminal speed.  As a result, the isovelocity zones,
as defined by the condition

\begin{equation}
v_{\rm z} = -v(r)\,\hat{r}\cdot\hat{z},
\end{equation}

\noindent are the same in our outflows as for a strictly spherical
wind with the same $\beta=1$ law.  The asymmetry arises entirely in
terms of density, which affects the optical depth and, in principle,
the source function.

The differential amount of luminosity contributed to the line emission
is given by

\begin{equation}
\Delta L_\nu = 4\pi \, j_\nu(r)\, {\cal P}_{\rm esc} \, \Delta V,
\end{equation}

\noindent where $j_\nu$ is the emissivity, ${\cal P}_{\rm esc}$
is an escape probability, and $\Delta V$ is a volume element in
the flow.  The emissivity is given by

\begin{equation}
j_\nu = \kappa_l\,\rho\,S_\nu\, \delta(\nu-\nu_{\rm z}),
\end{equation}

\noindent where $\kappa_l$ is the frequency integrated absorption
coefficient, $\rho$ the flow density, $S_\nu$ the source function,
and $\delta(x)$ the delta-function.  The argument of the delta function
represents the condition for the Sobolev approximation that the emission
appear at frequency $\nu$ where $\nu_{\rm z} = \nu_l\,(1+v_{\rm z}/c)$,
with $\nu_l$ the rest frequency of the line of interest.  The escape
probability is given by

\begin{equation}
{\cal P}_{\rm esc} = \frac{1-e^{\tau_S}}{\tau_S},
\end{equation}

\noindent where the Sobolev optical depth is

\begin{equation}
\tau_S = \frac{\kappa_l\,\rho\,\lambda_0}{(v/r)\,(1+\sigma\,\mu^2)},
\end{equation}

\noindent with $\sigma = -1 + d\ln v/d\ln r$.  Note that for expansion
at constant velocity, $\sigma=-1$, whereas for homologous expansion with
$v \propto r$, $\sigma =0$.

Using observer coordinates with radius $r$, azimuth $\alpha$ about the
observer axis, and $\mu$ the cosine of the polar angle from the observer
axis, the contribution to the line emission from a volume element becomes

\begin{equation}
\Delta L_\nu = 4\pi\,\kappa_l\,\rho\,S_\nu\,{\cal P}_{\rm esc}\,
	r^2\,\Delta r\,\Delta \alpha\,
	\left| \frac{\Delta \mu}{\Delta \nu_{\rm z}}\right|,
\end{equation}

\noindent where $\Delta \nu_{\rm z}/\Delta \mu = \nu_l\,\Delta v_{\rm z}/c
= -\nu_l\,v(r)/c$.

For our purposes, we consider modelling of optical recombination lines.
The source function is taken as a constant, and we parameterize the
opacity as

\begin{equation}
\kappa_l\,\rho \propto \rho^2\,w^{\alpha_1}\,(1-w)^{\alpha_2},
\end{equation}

\noindent which scales with the square of density, multiplied by two
functions that depend on the normalized velocity, $w=v(r)/v_\infty$.
\citet{1987ApJ...314..726L} in the introduction of their SEI method introduced
such functions as a means of accounting for ionization gradients  \citep[c.f. eqn (17) of][]{1994ApJ...435..416B}.   Here, the functions should be interpreted 
as convenient scalings that serve to shift the line opacity toward the inner or
outer wind.  Note that $\alpha_1=\alpha_2=0$ results in $\kappa_l\,\rho
\propto \rho^2$.  

\subsection{Line Profile Variability From our CIR Geometries}

From our simple time-dependant emission-line profiles, one result clearly emerged. All solutions with  i$_0$=166$^{\circ}$ and $\psi$=152$^{\circ}$ produce very little line-profile variability. In retrospect, this is not surprising. Indeed, recall that with these parameters characterizing the orientation of the stellar axis, we mainly need CIRs close to the poles to reproduce the polarization light curves (See Table 4).  As the star rotates, these CIRs do not change direction very much with respect to our line-of-sight (see Figure 3) and therefore they do not generate much spectroscopic variability.

For the other orientations of the stellar axis we found, i.e. i$_0$=166$^{\circ}$ and $\psi$=63$^{\circ}$, the results are more promising. Indeed in that configuration, we need CIRs in the equator to reproduce the polarization variability. This geometry now involves CIRs that are changing direction drastically with respect to the observer as the star rotates, going from pointing roughly towards the observer's line-of-sight to away from it. Although there are  no spectroscopic observations obtained simultaneously with the polarization, to our knowledge there has never been a report in the literature of non variable line profiles for this star. We therefore conclude that although the polarization observations are unable to distinguish between these two situations, the former one is excluded by the fact that it does not generate line profile variability. For this reason, we conclude that $\psi$=63$^{\circ}$ is a better solution and exclude the polarimetric fit with $\psi$=152$^{\circ}$.

We use the fitted parameters for Dataset 3 to illustrate our calculated line profile variations. In addition to the light from a spherical wind, we used for the first CIR, $\delta_1=0.3$,  $\phi_1 = 30^{\circ}$ and $\theta _1 = 89^{\circ}$ and for the second, $\delta_2 = 0.3$,  $\phi_2 = 299^{\circ}$ and $\theta _2 = 96^{\circ}$. Another question we need to address is that of the $\delta$ parameter ($=\eta \cos \beta (1-\cos ^2\beta)$). Although the linear polarization depends only on this parameter and is blind to the density contrast in the CIR, $\eta$, and the CIR opening angle, $\beta$, individually, the situation is different for spectroscopy. To calculate the variable line profiles, we need to include these parameters separately. We therefore present plots for different combinations of these parameters that correspond to the same value of $\delta$; $\beta = 15^{\circ}, 25^{\circ}, 35^{\circ}$ and $\eta = 4.64, 1.85, 1.11$ (i.e. $n_{CIR}/n_{sph}=5.64, 2.85, 2.11$). Figure 5 shows profile changes for these three combinations of $\eta$ and $\beta$ and for two different values of the photospheric radius ($R_{ph}=1.25 R_*$ and $R_{ph}=2.5 R_*$). Presented in the lower panels of each plot are a set of theoretical line profiles for phases covering one cycle of the rotation cycle superposed on the mean of these profile in red. In the top portion of each plot, we present a greyscale of the differences of these individual line profiles with the mean as a function of phase, as is often done for actual observations. 

One can clearly see in each case the non variable base profile from the spherical wind and the variable excess emission form the CIRs. As the density contrast is increased (and therefore the CIR half opening angle is decreased) the fraction of the total line profile from the CIR increases as well. However, even for a density contrast that is relatively small  ($\eta = 1.1$, i.e. $\sim$ a factor of two), the fraction of the line intensity from the excess CIR emission is more than 50\%, if we adopt a value of R$_{ph}$=2.5 R$_*$. Only if we decease this radius to 1.25R$_*$ with a density contrast of $\eta = 1.1$ are we able to reduce this fraction to $\sim$20\%, which is closer  to the observed value for this type of variability of around 5-10\% \citep{2011ApJ...736..140C}. Note that a value of $R_{ph}=1.25R_*$ is essentially the same as our adopted value listed in Table 3 and estimated independently from the observed characteristics of the wind ($R_*, \, \dot{M}$ and $v_{\infty }$). To obtain a level of variability that is even closer to the observed value with this value of $R_{ph}$, we would have to decrease the density contrast further and therefore increase the opening angle of the CIR. As for the pattern of variability, it is very similar to the observed one displaying a characteristic S-shape as a function of time. Note that the presence of a dark trace accompanying the bright trace is a consequence of the way the data are presented as these are differences with the mean and therefore for any given wavelength, the sum of all difference spectra must be zero.

\begin{figure}
\centering
\vskip -0.1truecm
\includegraphics[width=0.5\textwidth]{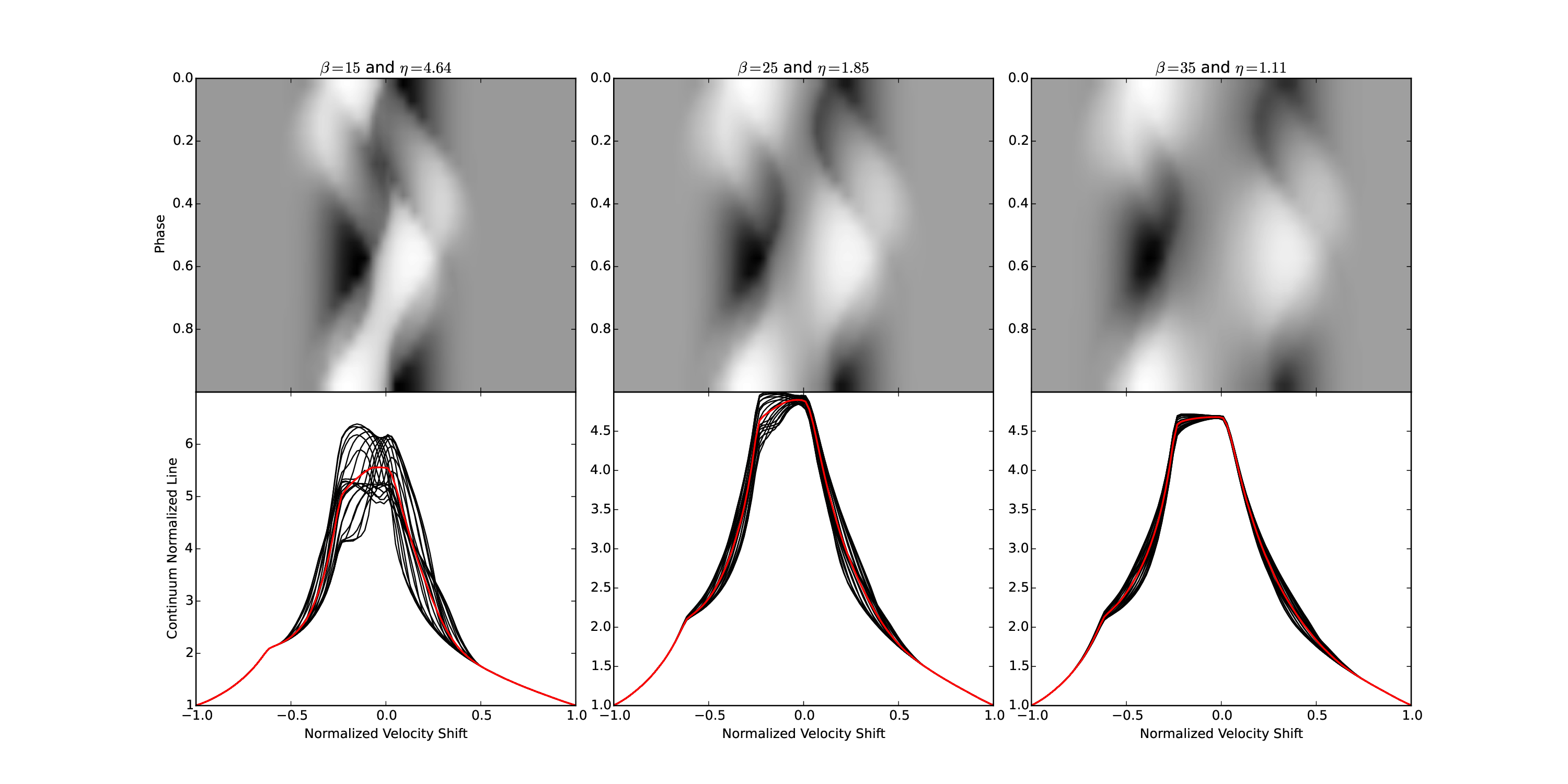}\\
\vskip -0.1truecm
\includegraphics[width=0.5\textwidth]{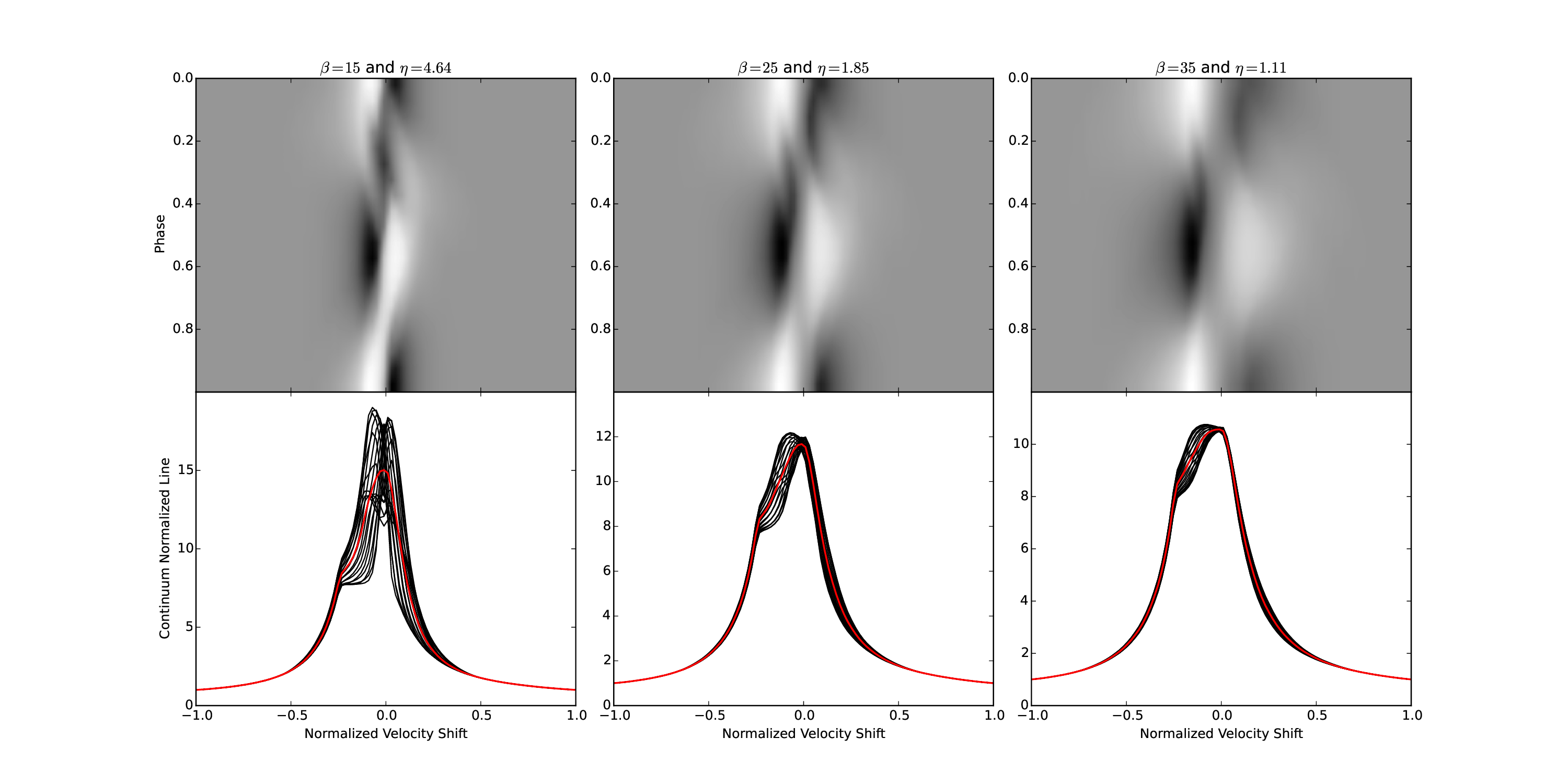}\\
\vskip -0.1truecm
\caption{Line profile variations for our polarimetric geometries using the fitted parameters of Dataset 3. Three combinations of $\beta$ and $\eta$ corresponding to $\delta=0.3$ for each CIR are shown. {\it Top:} for R$_{\tau=1}$=2.5 R$_*$. {\it Bottom:} for R$_{\tau=1}$=1.25 R$_*$.}
\end{figure}

\section{Discussion and Conclusion}
Our simplified model of time-dependant linear polarization from CIR-like structures in the spherical wind of a rotating massive star has allowed us to qualitatively reproduce all continuum polarization datasets of WR6 available in the literature with a consistent set of stellar parameters, i.e. an inclination of   $166 ^{\circ}$ and an orientation of the stellar axis in the plane of the sky of $\psi=63^{\circ}$.  From the polarimetry alone, this latter parameter was found to be ambiguous by $90^{\circ}$, that is two CIRs near the poles with $\psi=152^{\circ}$ produce very similar polarization signatures than two CIRs near the equator with $\psi=63^{\circ}$.  However, using Sobolev approximation line profile calculations, we were able to exclude the former solution as it leads to essentially no spectroscopic changes, which is incompatible with observations. 

With the above geometry, we find that including two CIRs in an otherwise spherical stellar wind is sufficient to reproduce the observed level of linear polarization as well as the general characteristics of the variability. For three datasets obtained over an uninterrupted 43-day period in February-April 1986, including three to six CIRs was found to improve the fits significantly. Interestingly, for our adopted value of $\psi = 63^{\circ}$, the two CIRs, in addition to being located near the stellar equator are always found to be separated in longitude by $\Delta \phi \approx 90^{\circ}$, but  to migrate slowly on the surface of the star with time, although the available data does not allow us to determine if this happens in any organized manner and on what exact timescale.  The density contrast of these spiral-like structures with the ambient wind and their opening angle cannot be fit individually using the polarization observations alone as these two parameters combine to affect the amplitude of the polarimetric variability. However, our Sobolev line profile simulations have helped to somewhat constrain these parameters. First, they show that for a photospheric radius, $R_{ph}$, that is too high, the contribution of the CIRs to the line profiles, which is variable, is far too strong compared to the constant contribution of the spherical wind, for density contrasts above about two.  For these density enhancements, much better agreement is obtained with a value of $R_{ph}$ around 1.25 $R_*$. Note however that this produces a level of variability that is about a factor of two too high. In order for our predicted variability levels to be closer to the observed ones,  the density contrast in the CIR must be smaller that a factor of two. This in turn indicates that the opening angles for the CIRs must be high, i.e. above $\beta \approx 35^{\circ}$. This is not incompatible with the values obtained from detailed 3D radiative transfer calculations and hydrodynamic models of CIRs in the wind of the early B supergiant star HD64760 by \citet{2008ApJ...678..408L}. The type of spectroscopic variability we predict with our simplified CIR model have, however, characteristics that are very reminiscent of that observed in the optical for this star.

Because the complete lack of absorption lines in the spectrum of WR6 excludes the presence of a massive O-type companion and the nature of the photometric (optical and X-ray), polarimetric and spectroscopic variability is unlikely to be caused by a low mass companion \citep[see discussion in][]{1997ApJ...482..470M}, the polarimetric fits presented in this work bring further support to a single-star interpretation of this star. In the case of CIRs, the period of the variations is directly linked to the rotation velocity of the star, which can be readily determined if the radius at which the CIRs originate is known. Because the surface of WR stars is impossible to observe directly because of their thick wind, very little is known about their rotation rates. Therefore, any diagnostic that can provide information on their value is valuable and can be used to constrain spin rates of massive-star end states. 

We have shown that the polarization variability is completely compatible with the spectroscopic variability if we interpret them in terms of the presence of CIRs in its wind.  Note that other types of large-scale wind asymmetries could possibly produce light curves with similar characteristics as the ones obtained here.  \citet{2000A&A...356..619B} have shown that density redistributions on a small scale such as the localised collapse of a wind region to form a small clump cannot lead to significant net polarization but that a large scale redistribution such as the collapse of an equatorial wedge into a disk could.  More detailed modelling would be required to verify if such structures are able to reproduce all observed polarization light curves with a coherent set of stellar parameters. If CIRs prove to be the correct interpretation,  the origin of the disturbance at the base of the wind that would give rise to them still remains unknown. However, the polarimetric modelling presented here provides some clues as to its property.  Indeed, within the framework of this interpretation, we find that two localized disturbances situated near the stellar equator and separated by $\sim 90^{\circ}$ in longitude were consistently found on the surface of the star for a period of at $\sim$5 years (from October 1985 to April 1990) and possibly up to 20 years (if one includes datasets 1 and 2), leading to large scale structures in the wind of this star. Since with this interpretation the modulation timescale is directly linked to the rotation velocity of the star, all that is required to explain the periodic albeit epoch-dependant variability is the presence of perturbations at the base of the wind at all times when variability is observed. The disappearance of the perturbations would most likely be followed by the dissipation of the CIR after a certain time but the emergence of other perturbations would lead to the creation of new CIRs, which would produce photometric, polarimetric and spectroscopic changes with different characteristics but with the same period.

Magnetic spots slowly drifting on the surface of the star are a possible candidate to produce such structures. Although an extensive spectropolarimetric dataset did not allow \citet{2013ApJ...764..171D} to detect a global dipolar magnetic field larger than $\sim$100 Gauss in the line forming region for this star, their data were not able to exclude the presence of smaller scale magnetic structures. To explain the epoch-dependancy of the periodic variability, such magnetic spots would have to have a finite lifetime of at least some forty days, as shown by our fits to datasets 4-6, and then fade away. In the absence of a perturbation, the spiral-like wind structure would then slowly dissipate. The appearance of more spots on other regions of the stellar surface would then subsequently produce new large-scale structures. The fact that we determined that the two base perturbations always seem to be distant by $\sim 90^{\circ}$ in longitude is certainly intriguing. 

A pattern of non-radial pulsations slowly moving along the star might also be a source of varying perturbations at the base of the wind. An uninterrupted light curve covering a timespan of the order of several months is the only means of searching for non-radial pulsation modes in the light of these stars. Very few massive-star pulsators are known, as because of the relatively long periods expected for these stars, extremely long, uninterrupted observing sequences are needed to identify the periods. For WR 6, only one period has been identified to date and it is found to be remarkably stable.

Of course, our model is simplified and lacks certain sophistications in terms of density, temperature and ionization structures, the presence of shocks, the velocity law within the CIRs, etc. Such complexification precludes an analytical prescription.  The value of an analytic approach is the ability to rapidly explore parameter space for achieving qualitative matches to observed behaviour.  Further progress in the quantitative modelling of CIRs in WR winds will likely be achieved through numerical simulations involving Monte-Carlo radiative transfer, which we are currently exploring.



\section*{Acknowledgements}

NSL acknowledges financial support from the Natural Science and Engineering Research Council (NSERC) of Canada.




\bibliographystyle{mnras}
\bibliography{st-louis} 

\begin{thebibliography}{}
\makeatletter
\relax
\def\mn@urlcharsother{\let\do\@makeother \do\$\do\&\do\#\do\^\do\_\do\%\do\~}
\def\mn@doi{\begingroup\mn@urlcharsother \@ifnextchar [ {\mn@doi@}
  {\mn@doi@[]}}
\def\mn@doi@[#1]#2{\def\@tempa{#1}\ifx\@tempa\@empty \href
  {http://dx.doi.org/#2} {doi:#2}\else \href {http://dx.doi.org/#2} {#1}\fi
  \endgroup}
\def\mn@eprint#1#2{\mn@eprint@#1:#2::\@nil}
\def\mn@eprint@arXiv#1{\href {http://arxiv.org/abs/#1} {{\tt arXiv:#1}}}
\def\mn@eprint@dblp#1{\href {http://dblp.uni-trier.de/rec/bibtex/#1.xml}
  {dblp:#1}}
\def\mn@eprint@#1:#2:#3:#4\@nil{\def\@tempa {#1}\def\@tempb {#2}\def\@tempc
  {#3}\ifx \@tempc \@empty \let \@tempc \@tempb \let \@tempb \@tempa \fi \ifx
  \@tempb \@empty \def\@tempb {arXiv}\fi \@ifundefined
  {mn@eprint@\@tempb}{\@tempb:\@tempc}{\expandafter \expandafter \csname
  mn@eprint@\@tempb\endcsname \expandafter{\@tempc}}}

\bibitem[\protect\citeauthoryear{{Aldoretta} et~al.,}{{Aldoretta}
  et~al.}{2016}]{2016MNRAS.460.3407A}
{Aldoretta} E.~J.,  et~al., 2016, \mn@doi [\mnras] {10.1093/mnras/stw1188},
  \href {http://adsabs.harvard.edu/abs/2016MNRAS.460.3407A} {460, 3407}

\bibitem[\protect\citeauthoryear{{Antokhin}, {Bertrand}, {Lamontagne}  \&
  {Moffat}}{{Antokhin} et~al.}{1994}]{1994AJ....107.2179A}
{Antokhin} I.,  {Bertrand} J.-F.,  {Lamontagne} R.,   {Moffat} A.~F.~J.,  1994,
  \mn@doi [\aj] {10.1086/117029}, \href
  {http://adsabs.harvard.edu/abs/1994AJ....107.2179A} {107, 2179}

\bibitem[\protect\citeauthoryear{{Bjorkman}, {Ignace}, {Tripp}  \&
  {Cassinelli}}{{Bjorkman} et~al.}{1994}]{1994ApJ...435..416B}
{Bjorkman} J.~E.,  {Ignace} R.,  {Tripp} T.~M.,   {Cassinelli} J.~P.,  1994,
  \mn@doi [\apj] {10.1086/174825}, \href
  {http://adsabs.harvard.edu/abs/1994ApJ...435..416B} {435, 416}

\bibitem[\protect\citeauthoryear{{Brown} \& {McLean}}{{Brown} \&
  {McLean}}{1977}]{1977A&A....57..141B}
{Brown} J.~C.,  {McLean} I.~S.,  1977, \aap, \href
  {http://adsabs.harvard.edu/abs/1977A%26A....57..141B} {57, 141}

\bibitem[\protect\citeauthoryear{{Brown}, {Carlaw}  \& {Cassinelli}}{{Brown}
  et~al.}{1989}]{1989ApJ...344..341B}
{Brown} J.~C.,  {Carlaw} V.~A.,   {Cassinelli} J.~P.,  1989, \mn@doi [\apj]
  {10.1086/167803}, \href {http://adsabs.harvard.edu/abs/1989ApJ...344..341B}
  {344, 341}

\bibitem[\protect\citeauthoryear{{Brown}, {Ignace}  \& {Cassinelli}}{{Brown}
  et~al.}{2000}]{2000A&A...356..619B}
{Brown} J.~C.,  {Ignace} R.,   {Cassinelli} J.~P.,  2000, \aap, \href
  {http://adsabs.harvard.edu/abs/2000A%26A...356..619B} {356, 619}

\bibitem[\protect\citeauthoryear{{Brown}, {Barrett}, {Oskinova}, {Owocki},
  {Hamann}, {de Jong}, {Kaper}  \& {Henrichs}}{{Brown}
  et~al.}{2004}]{2004A&A...413..959B}
{Brown} J.~C.,  {Barrett} R.~K.,  {Oskinova} L.~M.,  {Owocki} S.~P.,  {Hamann}
  W.-R.,  {de Jong} J.~A.,  {Kaper} L.,   {Henrichs} H.~F.,  2004, \mn@doi
  [\aap] {10.1051/0004-6361:20031557}, \href
  {http://adsabs.harvard.edu/abs/2004A%26A...413..959B} {413, 959}

\bibitem[\protect\citeauthoryear{{Cassinelli}, {Nordsieck}  \&
  {Murison}}{{Cassinelli} et~al.}{1987}]{1987ApJ...317..290C}
{Cassinelli} J.~P.,  {Nordsieck} K.~H.,   {Murison} M.~A.,  1987, \mn@doi
  [\apj] {10.1086/165277}, \href
  {http://adsabs.harvard.edu/abs/1987ApJ...317..290C} {317, 290}

\bibitem[\protect\citeauthoryear{{Chen{\'e}} \& {St-Louis}}{{Chen{\'e}} \&
  {St-Louis}}{2011}]{2011ApJ...736..140C}
{Chen{\'e}} A.-N.,  {St-Louis} N.,  2011, \mn@doi [\apj]
  {10.1088/0004-637X/736/2/140}, \href
  {http://adsabs.harvard.edu/abs/2011ApJ...736..140C} {736, 140}

\bibitem[\protect\citeauthoryear{{Cranmer} \& {Owocki}}{{Cranmer} \&
  {Owocki}}{1996}]{1996ApJ...462..469C}
{Cranmer} S.~R.,  {Owocki} S.~P.,  1996, \mn@doi [\apj] {10.1086/177166}, \href
  {http://adsabs.harvard.edu/abs/1996ApJ...462..469C} {462, 469}

\bibitem[\protect\citeauthoryear{{Dessart} \& {Chesneau}}{{Dessart} \&
  {Chesneau}}{2002}]{2002A&A...395..209D}
{Dessart} L.,  {Chesneau} O.,  2002, \mn@doi [\aap]
  {10.1051/0004-6361:20021110}, \href
  {http://adsabs.harvard.edu/abs/2002A%26A...395..209D} {395, 209}

\bibitem[\protect\citeauthoryear{{Drissen}, {Robert}, {Lamontagne}, {Moffat},
  {St-Louis}, {van Weeren}  \& {van Genderen}}{{Drissen}
  et~al.}{1989}]{1989ApJ...343..426D}
{Drissen} L.,  {Robert} C.,  {Lamontagne} R.,  {Moffat} A.~F.~J.,  {St-Louis}
  N.,  {van Weeren} N.,   {van Genderen} A.~M.,  1989, \mn@doi [\apj]
  {10.1086/167715}, \href {http://adsabs.harvard.edu/abs/1989ApJ...343..426D}
  {343, 426}

\bibitem[\protect\citeauthoryear{{Firmani}, {Koenigsberger}, {Bisiacchi},
  {Ruiz}  \& {Solar}}{{Firmani} et~al.}{1979}]{1979IAUS...83..421F}
{Firmani} C.,  {Koenigsberger} G.,  {Bisiacchi} G.~F.,  {Ruiz} E.,   {Solar}
  A.,  1979, in {Conti} P.~S.,  {De Loore} C.~W.~H.,  eds,  IAU Symposium Vol.
  83, Mass Loss and Evolution of O-Type Stars. pp 421--423

\bibitem[\protect\citeauthoryear{{Gayley} \& {Owocki}}{{Gayley} \&
  {Owocki}}{1995}]{1995ApJ...446..801G}
{Gayley} K.~G.,  {Owocki} S.~P.,  1995, \mn@doi [\apj] {10.1086/175837}, \href
  {http://adsabs.harvard.edu/abs/1995ApJ...446..801G} {446, 801}

\bibitem[\protect\citeauthoryear{{Gormaz-Matamala}, {Herv{\'e}}, {Chen{\'e}},
  {Cur{\'e}}  \& {Mennickent}}{{Gormaz-Matamala}
  et~al.}{2015}]{2015wrs..conf..357G}
{Gormaz-Matamala} A.~C.,  {Herv{\'e}} A.,  {Chen{\'e}} A.-N.,  {Cur{\'e}} M.,
  {Mennickent} R.~E.,  2015, in {Hamann} W.-R.,  {Sander} A.,   {Todt} H.,
  eds, Wolf-Rayet Stars: Proceedings of an International Workshop held in
  Potsdam, Germany, 1-5 June 2015. Edited by Wolf-Rainer Hamann, Andreas
  Sander, Helge Todt. Universit{\"a}tsverlag Potsdam, 2015., p.357. p.~357

\bibitem[\protect\citeauthoryear{{Hamann}, {Koesterke}  \&
  {Wessolowski}}{{Hamann} et~al.}{1995}]{1995A&A...299..151H}
{Hamann} W.-R.,  {Koesterke} L.,   {Wessolowski} U.,  1995, \aap, \href
  {http://adsabs.harvard.edu/abs/1995A%26A...299..151H} {299, 151}

\bibitem[\protect\citeauthoryear{{Hamann}, {Brown}, {Feldmeier}  \&
  {Oskinova}}{{Hamann} et~al.}{2001}]{2001A&A...378..946H}
{Hamann} W.-R.,  {Brown} J.~C.,  {Feldmeier} A.,   {Oskinova} L.~M.,  2001,
  \mn@doi [\aap] {10.1051/0004-6361:20011253}, \href
  {http://adsabs.harvard.edu/abs/2001A%26A...378..946H} {378, 946}

\bibitem[\protect\citeauthoryear{{Hamann}, {Gr{\"a}fener}  \&
  {Liermann}}{{Hamann} et~al.}{2006}]{2006AandA...457.1015H}
{Hamann} W.-R.,  {Gr{\"a}fener} G.,   {Liermann} A.,  2006, \mn@doi [\aap]
  {10.1051/0004-6361:20065052}, \href
  {http://adsabs.harvard.edu/abs/2006A%26A...457.1015H} {457, 1015}

\bibitem[\protect\citeauthoryear{{Howarth} \& {Prinja}}{{Howarth} \&
  {Prinja}}{1989}]{1989ApJS...69..527H}
{Howarth} I.~D.,  {Prinja} R.~K.,  1989, \mn@doi [\apjs] {10.1086/191321},
  \href {http://adsabs.harvard.edu/abs/1989ApJS...69..527H} {69, 527}

\bibitem[\protect\citeauthoryear{{Howarth} \& {Schmutz}}{{Howarth} \&
  {Schmutz}}{1995}]{1995AandA...294..529H}
{Howarth} I.~D.,  {Schmutz} W.,  1995, \aap, \href
  {http://adsabs.harvard.edu/abs/1995A%26A...294..529H} {294, 529}

\bibitem[\protect\citeauthoryear{{Huenemoerder} et~al.,}{{Huenemoerder}
  et~al.}{2015}]{2015ApJ...815...29H}
{Huenemoerder} D.~P.,  et~al., 2015, \mn@doi [\apj]
  {10.1088/0004-637X/815/1/29}, \href
  {http://adsabs.harvard.edu/abs/2015ApJ...815...29H} {815, 29}

\bibitem[\protect\citeauthoryear{{Ignace}, {Nordsieck}  \&
  {Cassinelli}}{{Ignace} et~al.}{2004}]{2004ApJ...609.1018I}
{Ignace} R.,  {Nordsieck} K.~H.,   {Cassinelli} J.~P.,  2004, \mn@doi [\apj]
  {10.1086/421258}, \href {http://adsabs.harvard.edu/abs/2004ApJ...609.1018I}
  {609, 1018}

\bibitem[\protect\citeauthoryear{{Ignace}, {Hubrig}  \&
  {Sch{\"o}ller}}{{Ignace} et~al.}{2009}]{2009AJ....137.3339I}
{Ignace} R.,  {Hubrig} S.,   {Sch{\"o}ller} M.,  2009, \mn@doi [\aj]
  {10.1088/0004-6256/137/2/3339}, \href
  {http://adsabs.harvard.edu/abs/2009AJ....137.3339I} {137, 3339}

\bibitem[\protect\citeauthoryear{{Ignace}, {St-Louis}  \&
  {Proulx-Giraldeau}}{{Ignace} et~al.}{2015}]{2015A&A...575A.129I}
{Ignace} R.,  {St-Louis} N.,   {Proulx-Giraldeau} F.,  2015, \mn@doi [\aap]
  {10.1051/0004-6361/201424806}, \href
  {http://adsabs.harvard.edu/abs/2015A%26A...575A.129I} {575, A129}

\bibitem[\protect\citeauthoryear{{Kaper}, {Henrichs}, {Nichols}, {Snoek},
  {Volten}  \& {Zwarthoed}}{{Kaper} et~al.}{1996}]{1996A&AS..116..257K}
{Kaper} L.,  {Henrichs} H.~F.,  {Nichols} J.~S.,  {Snoek} L.~C.,  {Volten} H.,
   {Zwarthoed} G.~A.~A.,  1996, \aaps, \href
  {http://adsabs.harvard.edu/abs/1996A%26AS..116..257K} {116, 257}

\bibitem[\protect\citeauthoryear{{Lamers}, {Cerruti-Sola}  \&
  {Perinotto}}{{Lamers} et~al.}{1987}]{1987ApJ...314..726L}
{Lamers} H.~J.~G.~L.~M.,  {Cerruti-Sola} M.,   {Perinotto} M.,  1987, \mn@doi
  [\apj] {10.1086/165100}, \href
  {http://adsabs.harvard.edu/abs/1987ApJ...314..726L} {314, 726}

\bibitem[\protect\citeauthoryear{{Lamontagne}, {Moffat}  \&
  {Lamarre}}{{Lamontagne} et~al.}{1986}]{1986AJ.....91..925L}
{Lamontagne} R.,  {Moffat} A.~F.~J.,   {Lamarre} A.,  1986, \mn@doi [\aj]
  {10.1086/114068}, \href {http://adsabs.harvard.edu/abs/1986AJ.....91..925L}
  {91, 925}

\bibitem[\protect\citeauthoryear{{Lobel} \& {Blomme}}{{Lobel} \&
  {Blomme}}{2008}]{2008ApJ...678..408L}
{Lobel} A.,  {Blomme} R.,  2008, \mn@doi [\apj] {10.1086/529129}, \href
  {http://adsabs.harvard.edu/abs/2008ApJ...678..408L} {678, 408}

\bibitem[\protect\citeauthoryear{{McLean}}{{McLean}}{1980}]{1980ApJ...236L.149M}
{McLean} I.~S.,  1980, \mn@doi [\apjl] {10.1086/183216}, \href
  {http://adsabs.harvard.edu/abs/1980ApJ...236L.149M} {236, L149}

\bibitem[\protect\citeauthoryear{{McLean}, {Coyne}, {Frecker}  \&
  {Serkowski}}{{McLean} et~al.}{1979}]{1979ApJ...231L.141M}
{McLean} I.~S.,  {Coyne} G.~V.,  {Frecker} S.~J.~J.~E.,   {Serkowski} K.,
  1979, \mn@doi [\apjl] {10.1086/183021}, \href
  {http://adsabs.harvard.edu/abs/1979ApJ...231L.141M} {231, L141}

\bibitem[\protect\citeauthoryear{{Moffat}, {Drissen}, {Lamontagne}  \&
  {Robert}}{{Moffat} et~al.}{1988}]{1988ApJ...334.1038M}
{Moffat} A.~F.~J.,  {Drissen} L.,  {Lamontagne} R.,   {Robert} C.,  1988,
  \mn@doi [\apj] {10.1086/166895}, \href
  {http://adsabs.harvard.edu/abs/1988ApJ...334.1038M} {334, 1038}

\bibitem[\protect\citeauthoryear{{Morel}, {St-Louis}  \& {Marchenko}}{{Morel}
  et~al.}{1997}]{1997ApJ...482..470M}
{Morel} T.,  {St-Louis} N.,   {Marchenko} S.~V.,  1997, \mn@doi [\apj]
  {10.1086/304122}, \href {http://adsabs.harvard.edu/abs/1997ApJ...482..470M}
  {482, 470}

\bibitem[\protect\citeauthoryear{{Morel} et~al.,}{{Morel}
  et~al.}{1999}]{1999ApJ...518..428M}
{Morel} T.,  et~al., 1999, \mn@doi [\apj] {10.1086/307250}, \href
  {http://adsabs.harvard.edu/abs/1999ApJ...518..428M} {518, 428}

\bibitem[\protect\citeauthoryear{{Nugis}, {Crowther}  \& {Willis}}{{Nugis}
  et~al.}{1998}]{1998AandA...333..956N}
{Nugis} T.,  {Crowther} P.~A.,   {Willis} A.~J.,  1998, \aap, \href
  {http://adsabs.harvard.edu/abs/1998A%26A...333..956N} {333, 956}

\bibitem[\protect\citeauthoryear{{Owocki} \& {Rybicki}}{{Owocki} \&
  {Rybicki}}{1984}]{1984ApJ...284..337O}
{Owocki} S.~P.,  {Rybicki} G.~B.,  1984, \mn@doi [\apj] {10.1086/162412}, \href
  {http://adsabs.harvard.edu/abs/1984ApJ...284..337O} {284, 337}

\bibitem[\protect\citeauthoryear{{Owocki}, {Castor}  \& {Rybicki}}{{Owocki}
  et~al.}{1988}]{1988ApJ...335..914O}
{Owocki} S.~P.,  {Castor} J.~I.,   {Rybicki} G.~B.,  1988, \mn@doi [\apj]
  {10.1086/166977}, \href {http://adsabs.harvard.edu/abs/1988ApJ...335..914O}
  {335, 914}

\bibitem[\protect\citeauthoryear{{Prinja} \& {Smith}}{{Prinja} \&
  {Smith}}{1992}]{1992A&A...266..377P}
{Prinja} R.~K.,  {Smith} L.~J.,  1992, \aap, \href
  {http://adsabs.harvard.edu/abs/1992A%26A...266..377P} {266, 377}

\bibitem[\protect\citeauthoryear{{Prinja}, {Barlow}  \& {Howarth}}{{Prinja}
  et~al.}{1990}]{1990ApJ...361..607P}
{Prinja} R.~K.,  {Barlow} M.~J.,   {Howarth} I.~D.,  1990, \mn@doi [\apj]
  {10.1086/169224}, \href {http://adsabs.harvard.edu/abs/1990ApJ...361..607P}
  {361, 607}

\bibitem[\protect\citeauthoryear{{Prinja} et~al.,}{{Prinja}
  et~al.}{1992}]{1992ApJ...390..266P}
{Prinja} R.~K.,  et~al., 1992, \mn@doi [\apj] {10.1086/171276}, \href
  {http://adsabs.harvard.edu/abs/1992ApJ...390..266P} {390, 266}

\bibitem[\protect\citeauthoryear{{Robert} \& {Moffat}}{{Robert} \&
  {Moffat}}{1989}]{1989ApJ...343..902R}
{Robert} C.,  {Moffat} A.~F.~J.,  1989, \mn@doi [\apj] {10.1086/167759}, \href
  {http://adsabs.harvard.edu/abs/1989ApJ...343..902R} {343, 902}

\bibitem[\protect\citeauthoryear{{Robert} et~al.,}{{Robert}
  et~al.}{1992}]{1992ApJ...397..277R}
{Robert} C.,  et~al., 1992, \mn@doi [\apj] {10.1086/171786}, \href
  {http://adsabs.harvard.edu/abs/1992ApJ...397..277R} {397, 277}

\bibitem[\protect\citeauthoryear{{Rybicki} \& {Hummer}}{{Rybicki} \&
  {Hummer}}{1983}]{1983ApJ...274..380R}
{Rybicki} G.~B.,  {Hummer} D.~G.,  1983, \mn@doi [\apj] {10.1086/161454}, \href
  {http://adsabs.harvard.edu/abs/1983ApJ...274..380R} {274, 380}

\bibitem[\protect\citeauthoryear{{Schulte-Ladbeck}, {Nordsieck}, {Nook},
  {Magalhaes}, {Taylor}, {Bjorkman}  \& {Anderson}}{{Schulte-Ladbeck}
  et~al.}{1990}]{1990ApJ...365L..19S}
{Schulte-Ladbeck} R.~E.,  {Nordsieck} K.~H.,  {Nook} M.~A.,  {Magalhaes} A.~M.,
   {Taylor} M.,  {Bjorkman} K.~S.,   {Anderson} C.~M.,  1990, \mn@doi [\apjl]
  {10.1086/185878}, \href {http://adsabs.harvard.edu/abs/1990ApJ...365L..19S}
  {365, L19}

\bibitem[\protect\citeauthoryear{{Schulte-Ladbeck}, {Nordsieck}, {Taylor},
  {Nook}, {Bjorkman}, {Magalhaes}  \& {Anderson}}{{Schulte-Ladbeck}
  et~al.}{1991}]{1991ApJ...382..301S}
{Schulte-Ladbeck} R.~E.,  {Nordsieck} K.~H.,  {Taylor} M.,  {Nook} M.~A.,
  {Bjorkman} K.~S.,  {Magalhaes} A.~M.,   {Anderson} C.~M.,  1991, \mn@doi
  [\apj] {10.1086/170717}, \href
  {http://adsabs.harvard.edu/abs/1991ApJ...382..301S} {382, 301}

\bibitem[\protect\citeauthoryear{{Serkowski}}{{Serkowski}}{1970}]{1970ApJ...160.1083S}
{Serkowski} K.,  1970, \mn@doi [\apj] {10.1086/150496}, \href
  {http://adsabs.harvard.edu/abs/1970ApJ...160.1083S} {160, 1083}

\bibitem[\protect\citeauthoryear{{Smith}, {Shara}  \& {Moffat}}{{Smith}
  et~al.}{1996}]{1996MNRAS.281..163S}
{Smith} L.~F.,  {Shara} M.~M.,   {Moffat} A.~F.~J.,  1996, \mnras, \href
  {http://adsabs.harvard.edu/abs/1996MNRAS.281..163S} {281, 163}

\bibitem[\protect\citeauthoryear{{Sobolev}}{{Sobolev}}{1960}]{1960mes..book.....S}
{Sobolev} V.~V.,  1960, {Moving envelopes of stars}

\bibitem[\protect\citeauthoryear{{St-Louis}, {Dalton}, {Marchenko}, {Moffat}
  \& {Willis}}{{St-Louis} et~al.}{1995}]{1995ApJ...452L..57S}
{St-Louis} N.,  {Dalton} M.~J.,  {Marchenko} S.~V.,  {Moffat} A.~F.~J.,
  {Willis} A.~J.,  1995, \mn@doi [\apjl] {10.1086/309706}, \href
  {http://adsabs.harvard.edu/abs/1995ApJ...452L..57S} {452, L57}

\bibitem[\protect\citeauthoryear{{Stenflo}}{{Stenflo}}{1994}]{1994ASSL..189.....S}
{Stenflo} J.,  ed. 1994, {Solar Magnetic Fields: Polarized Radiation
  Diagnostics}  Astrophysics and Space Science Library Vol. 189,
  \mn@doi{10.1007/978-94-015-8246-9.
}

\bibitem[\protect\citeauthoryear{{de la Chevroti{\`e}re}, {St-Louis}, {Moffat}
  \& {the MiMeS Collaboration}}{{de la Chevroti{\`e}re}
  et~al.}{2013}]{2013ApJ...764..171D}
{de la Chevroti{\`e}re} A.,  {St-Louis} N.,  {Moffat} A.~F.~J.,   {the MiMeS
  Collaboration} 2013, \mn@doi [\apj] {10.1088/0004-637X/764/2/171}, \href
  {http://adsabs.harvard.edu/abs/2013ApJ...764..171D} {764, 171}

\makeatother
\end{thebibliography}






\bsp	
\label{lastpage}
\end{document}